\documentclass{iopart}

\usepackage{fleqn}

\usepackage{ifthen}
\usepackage{ifpdf}

\usepackage{latexsym}
\usepackage{amssymb}
\usepackage{bm}

\ifpdf
\usepackage{graphicx}
\usepackage{epstopdf}
\else
\usepackage{graphicx}
\usepackage{epsfig}
\fi

\newcommand{\const}{\mbox{const}}

\newcommand{\im}{\mbox{Im}}

\newcommand{\mass}{\mathsf{m}}

\newcommand{\tbox}[1]{\mbox{\tiny #1}}

\newcommand{\be}[1]{\begin{eqnarray}\ifthenelse{#1=-1}{\nonumber}{\ifthenelse{#1=0}{}{\label{e#1}}}}
\newcommand{\ee}{\end{eqnarray}}

\newcommand{\hide}[1]{}
\newcommand{\Cn}[1]{\begin{center} #1 \end{center}}
\newcommand{\mpg}[2][\hsize]{\begin{minipage}[b]{#1}{#2}\end{minipage}}
\newcommand{\putgraph}[2][width=\hsize]{ \includegraphics[#1]{#2} }

\begin{document}


\title[Quantum pump in a closed circuit]
{Operating a quantum pump in a closed circuit}

\author{Itamar Sela and Doron Cohen}

\address{
Department of Physics, Ben-Gurion University, Beer-Sheva 84005, Israel
}


\begin{abstract}
During an adiabatic pumping cycle a conventional 
two barrier quantum device takes an electron 
from the left lead and ejects it to the right lead. 
Hence the pumped charge per cycle is naively 
expected to be $Q \le e$. This zero order adiabatic 
point of view is in fact misleading. For a closed device 
we can get ${Q > e}$ and even ${Q \gg e}$. 
In this paper a detailed analysis of the 
quantum pump operation is presented. 
Using the Kubo formula for the geometric 
conductance, and applying the Dirac chains picture,  
we derive practical estimates for~$Q$. 
\end{abstract}

\section{Introduction}

Understanding of charge transport in mesoscopic 
and molecular size devices is essential 
for the future realization of `quantum circuits' \cite{rev}.   
Of particular interest are adiabatic processes 
that take electrons and move them one by one 
via a device. The simple minded peristaltic point 
of view of such process is misleading:    
such picture looks valid in the case of an open circuit, 
but breaks down once the pump is integrated into 
a closed circuit \cite{pmp,pmc,pme}.
It is the purpose of this paper to further 
elaborate on the physics of quantum pumping 
in {\em closed circuits}, and to provide a detailed 
analysis of a prototype pump. 
The interest and the feasibility of realizing 
experiments with closed circuits is discussed 
in Section~1.3 of \cite{pml}.

\subsection{The prototype pumping device}

The prototype example for a quantum pumping 
device is the two barrier model (Fig.1). 
The one particle Hamiltonian is   
\be{1}
\mathcal{H}(X_1(t),X_2(t)) = \frac{1}{2\mass}\hat{p}^2 +
X_1(t)\delta(\hat{x}-x_1)+X_2(t)\delta(\hat{x}-x_2)
\ee
where $\mass$ is the mass of the particle 
and $(\hat{x},\hat{p})$ are the position 
and the momentum operators. 
The region $x<x_1$ is the left lead, 
and the region $x>x_2$ is the right lead.
We refer to the segment $x_1<x<x_2$ 
as the ``dot region". 
The driving is performed by 
changing $X_1$ and $X_2$ in time. 
In an actual experiment the control 
parameters $X_1$ and $X_2$ represent gate voltages.
In order to talk about charge transfer we 
have to define also a current operator. In what 
follows we use the conventional definition 
\be{2}
\mathcal{I}=\frac{e}{2\mass}
\left(\hat{p}\delta(\hat{x}-x_0) + \delta(\hat{x}-x_0)\hat{p}\right) 
\ee 
where $e$ is the charge of the particle 
and $x_0$ is an arbitrary section point. 
The momentary current via different sections 
is in general not the same. But if we integrate 
it over a whole pumping cycle the result becomes 
independent of $x_0$.

The pumping device can be used in 
two different configuration. 
In the case of an {\em open geometry} (Fig.1) 
the leads are attached to reservoirs 
that have the same chemical potentials.
For simplicity one assumes 
a zero temperature Fermi occupation.

In the case of a {\em closed geometry} (Fig.1)
the leads are detached from 
the reservoirs and are connected together 
so as to have a ring.    
This means periodic boundary conditions 
over a large space interval $(-L/2)<x<(L/2)$.
Furthermore, the closed system is assumed to be 
strictly isolated from any environmental influences. 
The closed system can be regarded 
as a network with two nodes that 
are connected by two bonds one 
of length $L_{\tbox{D}}$ (dot region) 
and the other of length $L_{\tbox{W}}$ (wire region). 
The total length of the ring 
is $\mathit{L}=L_{\tbox{D}}+L_{\tbox{W}}$.
We assume that $L_{\tbox{D}} \ll L$.

\subsection{Open (leads) geometry}

The {\em open version} of the two barrier model 
has been considered in Ref.\cite{tbm} 
using the scattering matrix formalism 
of B\"{u}ttiker Pr\'{e}tre and Thomas (BPT) \cite{BPT1,BPT2}. 
A typical pumping cycle is illustrated in Fig.2c.
In the 1st half of the cycle an electron 
is taken from the left lead into the dot region 
via the left barrier,  while in the second 
half of the cycle an electron is transfered 
from the dot region to the right lead  
via the right barrier. 
Naively, by this peristaltic picture, 
it seems that at most one electron 
is pumped through the device per cycle. 
This expectation is supported by the formal 
calculation. Using the BPT formula   
one obtains \cite{SAA}
\be{1000}
Q \ \ = \ \ (1-\overline{g}_{\tbox{T}}) e
\ee
where $0<\overline{g}_{\tbox{T}}<1$ characterizes 
the transmission of the device during the charge transfer. 
In the limit $\overline{g}_{\tbox{T}} \rightarrow 0$, which 
is a pump with {\em no leakage}, indeed one gets $Q=e$. 
Otherwise one gets $Q<e$.

\subsection{Closed (ring) geometry}

Our interest is in the {\em closed version} 
of the two barrier model. 
A major observation is that 
the pumped charge~$Q$ is {\em not} ``quantized" 
even if the device is closed and isolated from 
any environmental influences. 
Moreover, it can be larger than unit charge. 
In fact we can have $Q \gg e$.   

The analysis that we are going to present 
demonstrates and refines general results 
that were obtained in Ref.\cite{pmc}. 
There we have worked out an artificial circuit 
which has been modeled as a $3$~site system. 
In the present paper we would like to work out 
a major prototype model that allows 
the desired comparison between results 
for closed circuit as opposed to 
that of open geometry [Eq.(\ref{e1000})].

We are going to use the Kubo approach 
to quantum pumping \cite{pmp,pmc,pme}.    
The ``Dirac chains picture" 
which we further review in the next subsection   
makes a distinction between  
``near field" and ``far field" 
pumping cycles. The near field 
result has been further considered 
in Ref.\cite{pMB} using an extension 
of the BPT scattering approach to quantum pumping.

\subsection{The Dirac chains picture}

In order to analyze an {\em adiabatic} \cite{berry1} pumping cycle 
we have first to understand the geometry of the parameter space. 
In fact the parameter space of the two barrier model  
is three dimensional $(X_1,X_2,X_3)$ where $X_3=\Phi$ 
is the flux via the ring. In practice we assume 
a planar $\Phi=0$ pumping cycle, but for the 
theoretical discussion it is convenient 
to regard $\Phi$ as a free parameter. 

We ask what is the amount of charge which is transported 
via a section of the ring per cycle. For this 
purpose we have to calculate the 
current $I=\langle\mathcal{I}\rangle$ at each moment. 
If we were changing the flux we would have 
by Ohm law $I=-G^{33}\dot{\Phi}$ where $G^{33}$ 
is called the Ohmic (dissipative) conductance. 
But if we change (say) the parameter $X_1$ 
then $I=-G^{31}\dot{X_1}$, where  $G^{31}$ 
is called the geometric (non-dissipative) 
conductance \cite{thouless,AvronNet,berry2}.
In general we can write $dQ = -G^{31}dX_1 -G^{32}dX_2$ 
and hence     
\be{3}
Q \ = \ \oint I dt \ = \ \oint \bm{G} \cdot d\bm{X}
\ee 
where $\bm{X}=(X_1,X_2)$ and $\bm{G} = (G^{31},G^{32})$.

The elements of the conductance matrix $G^{kj}$ 
can be calculated using the Kubo formula. 
It turns out \cite{berry1,berry2} that in the adiabatic 
limit $G^{31}=B_2$  and $G^{32}=-B_1$ 
where $\vec{B}$ is the ``magnetic" field (2-form) 
which appears in the theory of Berry phase \cite{berry1}.  
The sources of this field are Dirac monopoles 
that are located at the points of degeneracy. 
For the double barrier model a given level $n$ 
can have a degeneracy provided $X_1=X_2$,  
and either $\Phi=0$ or $\Phi=\pi\hbar/e$ 
modulo $2\pi\hbar/e$.
In fact we have for each level (excluding the ground state) 
two Dirac chains of degeneracies as in Fig.2d.

From the above observation one easily 
draws the following conclusions:
\ \ {\bf (i)} We can get $Q \gg e$ for  
a tight cycle around a Dirac chain 
if the degeneracy is in the pumping plane. 
\ \ {\bf (ii)} We can get $Q \ll e$ for  
a tight cycle around a Dirac chain 
if the degeneracy is off the pumping plane. 
\ \ {\bf (iii)} We can get $Q \sim e$ for 
a cycle which is located in the far field 
of a Dirac chain. 
The existence of a far field region 
is not self evident. This constitutes 
a major motivation for the present study.

\subsection{outline}

The outline of this paper is as follows. 
In section~2 we clarify the starting point 
of the calculation, which is the Kubo formula.  
In section~3 we introduce a preliminary discussion 
of the expected results and their significance.
Then in sections~4 to~11 we analyze the 
pumping process in the two barrier model. 
In particular we find the dependence 
of~$Q$ on the ``radius" of the pumping cycle, 
and make a distinction between ``near field" 
and ``far field" results.

\section{The Kubo formula}

Given a time dependent 
Hamiltonian $\mathcal{H}(X)$ 
with $X=X(t)$ we define 
$\mathcal{F}=-\partial \mathcal{H}/\partial X$
and would like to calculate 
the generalized conductance 
as defined by the expression
\be{14}
\langle \mathcal{I} \rangle =
\ \ = \ \ - G \ \dot{X}
\ee
We label by $n$ the adiabatic levels 
of the closed ring. The adiabatic states  
are defined by the equation 
$\mathcal{H}|n\rangle = E_n |n\rangle$ 
with implicit $X$ dependence.
Using these notations the Kubo formula 
for the {\em geometric conductance}  
can be written as 
\be{15}
G = \sum_{m(\ne n)}
\frac{ 2\hbar\mathcal{F}_{mn} }
{(E_m-E_n)^2}\im[\mathcal{I}_{nm}]
\ee
The above expression assumes that only one 
level ($n$) is occupied. If several levels 
are occupied we have to sum over them.
If we have more than one control variable, 
say $X_1$ and $X_2$, then we have 
to use the more elaborated notation $G^{31}$ 
and $G^{32}$ in order to distinguish between 
different elements of the conductance matrix.

If the pumping cycle crosses very close 
to a degeneracy, we can get from Eqs.(\ref{e14}-\ref{e15})
a very large current $I$, and upon integration 
we can find that the transported charge is $Q\gg e$. 
In the next sections we shall develop actual 
estimates for $Q$. But first we would like 
to further illuminate the significance of~$Q$.

The Schrodinger equation can be written in the {\em adiabatic basis}, 
where the {\em transformed} Hamiltonian matrix takes the form 
$H_{nm} = E_n \delta_{nm}  + \dot{X} A_{nm}$, 
where $E_n$ are called the adiabatic 
energy levels, and $A_{nm}$ is a matrix 
that can be calculated using a well known recipe 
(which is summarized in section~III of \cite{pmc}).   
One regards $\dot{X}$ as the ``small parameter". 
If the system is prepared an 
instantaneous {\em adiabatic} state $|n\rangle$,  
then the instantaneous current is $I=0$, 
whereas if it is prepared in an  
instantaneous {\em steady} state 
(an eigenstate of $H_{nm}$),   
then the instantaneous current is finite, say $I=I_0$.
Accordingly one can question the physical 
relevance of~$Q$: After all typically 
the initial preparation is an adiabatic state, 
and not a steady state.

So let us consider an actual physical scenario. 
For simplicity we assume that only two adiabatic 
levels are involved: The occupied level $n$ 
and next (empty) level $m=n+1$. 
To make the dynamical picture simple we use 
an analogy with the dynamics of a spin $1/2$ particle,  
and consider the illustration in Fig.3.  
We regard the state $n$ as spin polarized 
in the~$z$ direction, and the state $m$
as spin polarized in the $-z$ direction. 
The instantaneous steady states of the Hamiltonian 
are polarized along an axis that has a small tilt 
relative to the~$z$~direction.

Initially the spin is polarized in the $z$ direction 
and therefore $I(t=0)=0$. 
For some time we have $|X(t)-X(0)| \ll \delta X_c$ 
where $\delta X_c$ is the relevant parametric scale 
for the variation of the adiabatic levels.
During this time interval the tilt angle is 
approximately constant. The spin is performing 
a precession around the tilted axis. 
As a results we get $I(t) \ne 0$. 
In fact the maximum current is $I(t)=2I_0$.  
We get this current after half period of precession.

So we have a modulated current $I(t)$ that equals 
upon averaging $I_0$. As long as $|X(t)-X(0)| \ll \delta X_c$ 
the precession goes on as described above. But 
on larger time scales we have to take into account 
the variation in the tilt angle. Consequently the 
modulation of the current is no longer in the 
range $0 < I(t) < 2I_0$, but rather it is shifted 
and may increase. Still the average stays approximately $I_0$.

Thus we see that in the actual physical scenario 
the average over $I(t)$ is the same as 
the $I_0$ of the instantaneous steady state. 
The validity conditions for this statement 
are essentially the validity conditions of 
linear response theory, which are further explained  
in Ref.\cite{pmc}. The discussion above illuminates 
the justification for the use of the first order 
steady state solution of Kubo for the purpose of evaluating 
the pumped charge in an actual physical scenario.

\section{Charge transfer during an avoided crossing}

The pumped charge $Q$ is obtained via the integral Eq.(\ref{e3}) 
using Eq.(\ref{e15}) for $G$. On the basis of the naive heuristic 
picture of the Introduction (and see Fig.2) we expect 
that most of the contribution to the integral would 
come from a small segments in $\bm{X}$ space where the {\em last} 
occupied level has an avoided crossing with the first 
unoccupied level. Later we define precisely 
the region in $\bm{X}$ space where this assumption 
is a valid approximation. 

Let us try to sketch what might come out from the 
calculation. Later on we are going to do a proper job. 
But before we dive into the detailed analysis 
(which is quite lengthy) it would be nice to gain 
some rough expectation.  
Given that our interest is focused 
in a small energy window such that $E_n \sim E$ 
we define $v_{\tbox{E}}=(2E/\mass)^{1/2}$. 
The mean level spacing in the energy 
range of interest is  
\be{0}
\Delta = v_{\tbox{E}} \frac{\hbar\pi}{L}
\ee 
while the energy splitting $\Delta_s$  
at the avoided crossing might be much smaller.
We define the following notations
\be{0}
|\mathcal{F}_{nm}| &\equiv& \sigma_0 \\
|\mathcal{I}_{nm}| &\equiv& \frac{ev_{\tbox{E}}}{L} \sqrt{g_{\varphi}} \\ 
\Delta_s/\Delta &\equiv& \sqrt{1-g_0}
\ee
where both $0<g_{\varphi}<1$ and $0<g_0<1$ 
are dimensionless. Note that $g_0$ is related 
to the overall transmission $\overline{g}_{\tbox{T}}$ 
of the device. The adiabaticity condition is 
\be{0}
|\dot{X}| \ll \frac{\Delta_s^2}{\hbar\sigma_0} 
\ee
and from Eq.(\ref{e14}) with (\ref{e15}) we get the current
\be{0}
\langle \mathcal{I} \rangle  = 
2\left(\frac{\hbar\sigma_0}{\Delta_s^2}\dot{X}\right) 
\left(\frac{ev_{\tbox{E}}}{L}\right) \sqrt{g_{\varphi}}
\ee
The time of the Landau-Zener transition 
via the avoided crossing is  
\be{0}
\delta t \approx (\Delta_0/\sigma_0)/\dot{X}
\ee
and hence the transported charge is 
\be{23}
Q  
\ \ \approx \ \   
\langle \mathcal{I} \rangle \delta t 
\ \ \approx \ \  
\left(\frac{g_{\varphi}}{1-g_0}\right)^{1/2} e
\ee
where $g_0$ and $g_{\varphi}$ should be estimated 
in the region of the avoided crossing.  
We note that $g_{\varphi}/(1-g_0)$ 
is like the Thouless conductance, 
and can be regarded as a measure for 
the sensitivity of the energy levels 
to a test flux. We have pointed out 
and discussed this issue in Refs.\cite{pmp,pmc}, 
and later it was derived \cite{pMB} in the context 
of the scattering formalism.

In Fig.4 we display the numerically determined $Q$ 
for various path segments. The horizontal axis is 
the $|X_1-X_2|$ distance of the path segment from 
the degeneracy point. As $|X_1-X_2|$ becomes small 
$g_0 \rightarrow 1$  and we get $Q \gg e$. 
But the asymptotic value $Q \approx e$ which is 
observed for large $|X_1-X_2|$ cannot be explained 
by such a simple minded calculation. 
A major objective of the detailed analysis 
is to illuminate the observed crossover.

\section{The model, basic equations}

The one particle Hamiltonian of the two barriers model
depends on set of parameters $(X_1, X_2, \Phi)$.
From now on we use units such that $e=\hbar=1$, and    
characterize the geometry by the dimensionless parameter 
\be{0}
b \ \ = \ \ L_{\tbox{W}} / L_{\tbox{D}} \ \ \gg \ \ 1
\ee
We write the wavefunction on the two bonds as: 
\be{0}
\psi_{\tbox{dot}}(x)
&=& 
\left({2}/{\mathit{L}}\right)^{1/2} 
\sqrt{q_{\tbox{D}}} 
\ \sin({\varphi_{\tbox{D}}(x))}
\\
\psi_{\tbox{wire}}(x)
&=& 
\left({2}/{\mathit{L}}\right)^{1/2} 
\sqrt{q_{\tbox{W}}} 
\ \sin({\varphi_{\tbox{W}}(x))}
\ee
where $\varphi(x) = k x + \const$. 
Given $(X_1,X_2)$ and assuming $\Phi=0$,    
the eigenstates of this Hamiltonian can be found by 
searching $k_n$ values for which the 
matching conditions on the log-derivatives 
are satisfied. This leads to the following 
system of equations:
\be{27}
&& \cot(\varphi_{\tbox{W1}}) + \cot(\varphi_{\tbox{D1}}) = \frac{2\mass}{k} X_1 
\\ \label{e28}
\label{MatchingEquation2}
&& \cot(\varphi_{\tbox{W2}}) + \cot(\varphi_{\tbox{D2}}) = \frac{2\mass}{k} X_2 
\\
&& \varphi_{\tbox{D2}} - \varphi_{\tbox{D1}} = kL_{\tbox{D}} 
\\
&& \varphi_{\tbox{W2}} - \varphi_{\tbox{W1}} = kL_{\tbox{W}}
\ee
where it should be clear 
that $\varphi_{\tbox{D1}} \equiv \varphi_{\tbox{D}}(x_1)$ etc.
The corresponding eigenenergies are $E_n=k_n^2/2\mass$. 
A similar system of equation can be written for $\Phi=\pi$.
We can find the $q_{\tbox{W}}/q_{\tbox{D}}$ ratio for 
a given eigenstate via the matching condition 
on the wavefunction at either of the two nodes:
\be{31}
\sqrt{q_{\tbox{W}}} \ \sin({\varphi_{\tbox{W}})} 
= \sqrt{q_{\tbox{D}}} \ \sin({\varphi_{\tbox{D}})} 
\ee
where ${(\varphi_{\tbox{D}}, \varphi_{\tbox{W}})}$ mean 
either ${(\varphi_{\tbox{D1}},\varphi_{\tbox{W1}})}$
or ${(\varphi_{\tbox{D2}}, \varphi_{\tbox{W2}})}$. 
The normalization condition implies that 
\be{32}
\frac{L_{\tbox{D}}}{\mathit{L}} q_{\tbox{D}} 
+ \frac{L_{\tbox{W}}}{\mathit{L}} q_{\tbox{W}} 
\ \approx \ 1
\ee
For an ``ergodic state" we have $q\approx1$ for both bonds. 
In general we characterize an eigenstate by a the mixing parameter 
\be{33}
\Theta  
\equiv
2 \arctan\left(  \sqrt{ \frac{\mbox{Prob(wire)}}{\mbox{Prob(dot)}} }\right)    
=  2 \arctan\left( \sqrt{b}
\left(\frac{q_{\tbox{W}}}{q_{\tbox{D}}}\right)^{1/2}  \right)
\ee
such that $\Theta=0$ means a definite dot state, 
while $\Theta=\pi$ means a definite wire state. 
In practice we have ${0<\Theta<\pi}$.

In the numerical analysis we use units 
such that $\mass=1$ and $L_{\tbox{D}}=1$. 
Given $X_1$ and $X_2$ we solve the system 
of equations for the $k_n$ and for 
the $\varphi^{(n)}$ at the nodes. 
Then we determine $q^{(n)}$ 
at each bond, and from the ratio we 
get $\Theta^{(n)}$ as well.

\section{Regions in $\bm{X}$ space}

We focus on a small energy window 
such that the energy levels of interest 
are $E < E_n < E+dE$. We characterize 
a barrier $X\delta()$ using its transmission 
\be{0}
g(X) = \left[ 1 + \left(\frac{1}
{v_{\tbox{E}}}X \right)^2 \right]^{-1} 
\approx 
\left(\frac{1} {v_{\tbox{E}}}X \right)^{-2}
\ee  
The last expression holds if $g(X) \ll 1$.
We can regard $g_1=g(X_1)$ and $g_2=g(X_2)$ 
as an alternate way to specify $X_1$ and $X_2$.  
The $(X_1,X_2,\Phi)$ space is divided 
into various regions (Fig.5a). There is a region 
where $g_1$ and $g_2$ are of order one 
($|1-g| \ll 1$). There the delta functions 
at the nodes can be treated as a small perturbation. 
There is a region where ${1/b \ll g_1,g_2 \ll 1}$. 
There each dot level mix with many wire 
levels. This intermediate region will 
be discussed in a future work \cite{pmm}. 
Finally there is the region in $\bm{X}$ space 
which is of interest in the present study:
\be{34}
g_1,g_2 \ \ \ll \ \ 1/b  
\ee
We shall argue that in this region 
the states are categorized into 
``wire states" and ``dot state", 
which mix only whenever the energy 
level of the dot ``cross" 
an energy level of the wire (Fig.5bcd). 
This allows to use ``two level" 
approximation in the analysis of the mixing.

The degeneracies of the Hamiltonian occur 
at points $(X^{(r)},X^{(r)})$ along the
symmetry axis of $\bm{X}$ space. They are divided 
into two groups: those that are located 
in the $\Phi=0 (\mbox{mod}(2\pi))$ planes and those 
that are located in the $\Phi=\pi(\mbox{mod}(2\pi))$ planes.
In Appendix~A we find the explicit expression 
for $X^{(r)}$. It should be clear that each degeneracy 
point is duplicated $\mbox{mod}(2\pi)$ 
in the $\Phi$ direction, hence creating 
what we call a ``Dirac chain".

There are only two non-trivial  
degeneracies in $\bm{X}$ space which are 
associated with a given level $n$. 
One is with the neighboring level 
``from above" and the other is with the 
neighboring level ``from below".

Once we locate a degeneracy point  
we can make a distinction between ``near field" 
and ``far field" regions. 
The near field is defined as the region 
where we can use degenerate perturbation 
theory in order to figure out the 
splitting of the levels. 
In contrast to that, 
the far field is defined as the region 
where we can calculate the 
splitting of the levels 
by treating the dot-wire coupling 
as a first order perturbation.

In Fig.5a we show one pumping path that 
crosses in the near field region, 
and a second pumping path that 
crosses in the far field region.

\section{Eigenstate analysis}

In the theoretical analysis it is illuminating 
to map the behavior of $k_n$ and $\Theta^{(n)}$ 
in $(X_1,X_2,\Phi)$ space. See Figs.6-7.
It is not difficult to realize that the 
variation of $\Theta$ is bounded as follows:
\be{35}
{\sqrt{b}}\left[\frac{1}{4}g\right]^{+1/2}   
\ \ < \ \ \tan(\Theta/2) \ \ < \ \  
{\sqrt{b}}\left[\frac{1}{4}g\right]^{-1/2} 
\ee
where $g$ is either $g_1$ or $g_2$.  
The derivation of this result is as follows: 
The matching conditions at a given node 
implies that $(\varphi_{\tbox{D}}, \varphi_{\tbox{W}})$ 
are constraint to be on one of two 
branches which are illustrated in Fig.8. 
With each point of a given branch we can 
associate a $\Theta$ value via Eq.(\ref{e33})
and either Eq.(\ref{e27}) or Eq.(\ref{e28}). 
It is a straightforward exercise to 
express $\Theta$ say as a function 
of $\varphi_{\tbox{D}}$, and then to find 
its minimum and maximum values. Assuming 
that $g\ll 1$ one obtains Eq.(\ref{e35}).

Given $\Theta^{(n)}$ we can extract 
what are the $q^{(n)}$ at each bond, 
and what are the $\varphi^{(n)}$ at the nodes. 
By solving Eq.(\ref{e32}) with Eq.(\ref{e33}) we get  
\be{36}
q_{\tbox{D}} &\ \ = \ \ & b \ (\cos(\Theta/2))^2
\\
\label{e37}
q_{\tbox{W}} &\ \ = \ \ & (\sin(\Theta/2))^2
\ee
then from Eq.(\ref{e31}) 
with either Eq.(\ref{e27}) or Eq.(\ref{e28}) 
we find an expression for $\varphi$ at a given node. 
In particular we get 
\be{38}
\varphi_{\tbox{D}} = 
\left( 1 \pm \frac{1}{\sqrt{b}}\tan(\Theta/2) + ... \right) 
\left[\frac{1}{4}g\right]^{1/2} 
\ee
Note that $\varphi$ is well defined $\mbox{mod}(\pi)$. 
There are states which are dot-like ($\Theta \ll 1$),   
and there are states which are wire-like ($\Theta \sim \pi$). 
There are regions where a dot-like state mix with 
wire-like state leading to pair of states 
with $\Theta^{(+)} \sim \Theta^{(-)} \sim \pi/2$.     
From the above formula it follows that in the latter case 
\be{39}
&&\varphi_{\tbox{D}}^{(+)} 
\ \ \approx \ \ \varphi_{\tbox{D}}^{(-)} 
\ \ \approx \ \ \left[\frac{1}{4}g\right]^{1/2} 
\\ \label{e40}
\label{varphi_D_Minus_varphi_D}
&&|\varphi_{\tbox{D}}^{(+)} - \varphi_{\tbox{D}}^{(-)}| 
\ \ = \ \ 
\frac{1}{\sqrt{b}} \left[\frac{1}{4}g\right]^{1/2}
\left( \tan(\Theta^{(+)}/2)   + 
\tan(\Theta^{(-)}/2) \right)
\ee

\section{Eigenenergies and "mixing" from perturbation theory}

In order to find the splitting and the mixing of the two levels 
we use perturbation theory once in the far field analysis 
and once in the near field analysis. 
In both cases we use the following notations.  
The unperturbed basis is $\vert i \rangle$ with a dot-like state 
$|1\rangle$ and a wire-like $|2\rangle$. 
The perturbed eigenstates $\vert n \rangle$  
are indicated as $|+\rangle$ and $|-\rangle$. 
The Hamiltonian in both cases has the general form
\be{0}
\mathcal{H}=\mathcal{H}_0+\bm{W}=\left(
\begin{array}{cc}
E_1 & 0   \\
0   & E_2
\end{array}
\right)
+\left(
\begin{array}{cc}
W_{11} & W_{12}   \\
W_{21} & W_{22}
\end{array}
\right)
\ee
The eigenvectors are real so $W_{12}=W_{21}$.
The Hamiltonian can be written 
as a linear combination of Pauli matrices
\be{0}
\mathcal{H}=\left(\frac{E_1 + E_2}{2} + \frac{W_{11}+W_{22}}{2}\right){\bm 1}
+\left( \frac{E_1 - E_2}{2} + \frac{W_{11}-W_{22}}{2} \right)
{\bm \sigma}_3+W_{12}{\bm \sigma}_1
\ee
we define
\be{43}
\Delta_s &=& 2\sqrt{\left( \frac{E_1 - E_2}{2} + \frac{W_{11}-W_{22}}{2} \right)^2 + W_{12}^2} 
\\
\label{e44}
\tan(\theta) &=& \frac{2W_{12}}{(W_{11}-W_{22}) + (E_1 - E_2)}
\ee
The eigenenergies are
\be{45}
E_n =
\left(\frac{E_1 + E_2}{2} + \frac{W_{11}+W_{22}}{2}\right) 
\pm\frac{\Delta_s}{2} 
\ee
The eigenstates are found by rotating 
a spin half around the $y$ axis at an angle~$\theta$
\be{0}
|+\rangle
&\longrightarrow&
\left( \begin{array}{c} 
\cos{(\theta/2)} \\ \sin{(\theta/2)}
\end{array} \right)\\
\label{nVector}
|-\rangle
&\longrightarrow&
\left( \begin{array}{c} 
-\sin{(\theta/2)} \\ \cos{(\theta/2)}
\end{array} \right)
\ee
Assuming that the unperturbed basis 
consists of distinct dot-like 
and wire-like states, it follows that    
we can use the following approximation:
\be{48}
\Theta^{(+)} \ \ &=& \ \ \theta \\
\label{e49}
\Theta^{(-)} \ \ &=& \ \ \pi-\theta
\ee
Using Eq.(\ref{e36}) and Eq.(\ref{e40})
this implies
\be{50}
\sqrt{q_{\tbox{D}}^{(+)} q_{\tbox{D}}^{(-)}} 
\ \ &\approx& \ \ \frac{b}{2}|\sin(\theta)| 
\\ \label{e51}
|\varphi_{\tbox{D}}^{(+)} - \varphi_{\tbox{D}}^{(-)}|
\ \ &=& \ \ \frac{2}{\sqrt{b}} \left[\frac{1}{4}g\right]^{1/2}
\frac{1}{|\sin(\theta)|}
\ee
We note that in the strong mixing region we have $\theta\approx\pi/2$. 

In the next sections we obtain explicit expressions 
for the ``splitting" and the ``mixing" using the above 
scheme. From the numerics (see e.g. Fig.7) we see
that these are in fact very satisfactory approximations.

\section{Far field analysis}

The eigenstates of the ring in the $\bm{X}$ region 
of interest as defined in Eq.(\ref{e34}) can be found 
using first order perturbation theory with respect 
to the wire-dot coupling. This is explained below. 
We have verified numerically that the approximation 
is remarkable unless we are very close to 
a degeneracy point. This defines our distinction 
between ``far" and ``near" field regions. 
In the latter case we present in the next section 
a complementary treatment using degenerate 
perturbation theory.  
 
The unperturbed Hamiltonian in the far field 
analysis corresponds to ${X_1=X_2=\infty}$. 
Using the notations as defined in the previous  
section we take $|2\rangle$ as the $n^{\tbox{th}}$ 
wire state, and $|1\rangle$ as the closest 
dot state from above. Hence 
\be{52}
E_1 &=&  \frac{1}{2\mass} \left(\frac{\pi}{L_{\tbox{D}}}(1+[n/b]_{\tbox{integer}})\right)^2  \\
\label{e53}
E_2 &=&  \frac{1}{2\mass} \left(\frac{\pi}{L_{\tbox{W}}}n\right)^2
\ee
Using the formula
\be{0}
W_{ij} \ \ = \ \ -\frac{1}{4\mass^2 X} [\partial \psi^{(i)}] [\partial \psi^{(j)}] 
\ee
we get:
\be{0}
W_{11} &=& -\frac{v_{\tbox{E}}^2}{2L_{\tbox{D}}} 
\ \left[\frac{1}{X_1} + \frac{1}{X_2} \right] \\
W_{22} &=& -\frac{v_{\tbox{E}}^2}{2L_{\tbox{W}}}
\ \left[\frac{1}{X_1} + \frac{1}{X_2} \right] \\
\label{e57}
|W_{12}| &=& \frac{v_{\tbox{E}}^2}{2\sqrt{L_{\tbox{D}} L_{\tbox{W}}}} 
\ \left|\frac{1}{X_1} \pm \frac{1}{X_2} \right| 
\ee
where the $\pm$ sign in the expression for the dot-wire 
coupling depends on whether the dot and the wire states 
have the same parity or not. 

In case of a far field pumping cycle we start 
(say) with very high barriers, and then lower 
one of them, say $X_1$. If we neglected  
the dot-wire coupling $W_{12}$, the dot level   
would cross the wire level at a point $X_1=X^{(n)}$ 
that can be determined from the equation $E_1+W_{11}=E_2$.
At the vicinity of the avoided crossing we obtain 
\be{0} 
|W_{12}| = \frac{1}{2\pi} (b g^{(n)})^{1/2} \Delta 
\ee
where
\be{0}
g^{(n)} = \left(\frac{1}{v_{\tbox{E}}}X^{(n)}\right)^{-2} 
\ee
From the condition $|W_{12}|\ll \Delta$  
we deduce Eq.(\ref{e34}) which defines 
our $\bm{X}$ region of interest. 
Furthermore, from the results of the previous 
section we obtain expressions for the 
splitting and for the mixing of the levels:  
\be{60}
\Delta_s  &=&  bg^{(n)} \ \frac{1}{2L} ||\bm{X}-\bm{X}^{(r)}|| \\
\sin(\theta)  &=&  \frac{2}{\sqrt{b g^{(n)})}} 
\ \frac{v_{\tbox{E}}}{||\bm{X}-\bm{X}^{(r)}||}
\ee
We use the following notation, which 
we regard as a measure for the distance 
of the pumping cycle from the nearby degeneracy: 
\be{0}
||\bm{X}-\bm{X}^{(r)}||  = \sqrt{(X_1 - X^{(n)})^2 
+ \frac{4}{b}\left(\frac{v_{\tbox{E}}}{\sqrt{g^{(n)}}}\right)^2}
\ee
The significance of this notation   
will be further clarified in the next 
section where we extend the analysis 
into the near field region.

\section{Near field analysis}

The unperturbed Hamiltonian in the near field 
analysis corresponds to ${X_1=X_2=X^{(r)}}$. 
We can find a rough approximation for $X^{(r)}$ 
using the analysis of the previous section, 
but in fact we can find the exact expression  
which is derived in Appendix~A, 
where we also define the obvious 
notations $k_r$ and $g^{(r)}$. 
For later calculation the following 
approximation is useful: 
\be{0}
X^{(r)} \ \ \approx \ \ \frac{v_{\tbox{E}}}{\sqrt{g^{(r)}}}
\ee
The unperturbed basis consists of 
the dot-like definite 
parity state $|1\rangle$, 
and the wire-like definite parity state $|2\rangle$. 
We recall that for these states $\Theta$ 
attains its extremal values as remarked 
at the end of Appendix~A. 
The energies of the unperturbed states are
\be{0}
E_1 = E_2 = E_r = \frac{1}{2\mass}k_r^2
\ee
and the perturbation matrix is    
\be{0}
W_{11} &=& bg^{(r)} \frac{1}{2L} \left( \delta X_1 + \delta X_2 \right)\\
W_{22} &=& g^{(r)} \frac{1}{2L} \left( \delta X_1 + \delta X_2 \right)\\
|W_{12}| &=& \sqrt{b}g^{(r)} \frac{1}{2L} | \delta X_1 - \delta X_2 |
\ee
where
\be{0}
\delta X_1 &=& X_1 - X^{(r)} \\
\delta X_2 &=& X_2 - X^{(r)} 
\ee
Consequently we can determine both the 
splitting and the mixing of the levels:
\be{0}
\Delta_s  &=&  bg^{(r)}\frac{1}{2L} ||\bm{X}-\bm{X}^{(r)}|| \\
\sin(\theta)  &=& 
\frac{2}{\sqrt{b}} \  
\frac{|X_1-X_2|}{||\bm{X}-\bm{X}^{(r)}||}
\ee
In the above expression we have extended 
the interpretation of the distance measure 
as follows:
\be{0}
||\bm{X}-\bm{X}^{(r)}||  =
\sqrt{
\left( {X_1+X_2}-2X^{(r)}\right)^2 +
\frac{4}{b} |X_1- X_2|^2}
\ee
Now we realize that the far field results of the 
previous section are formally a saturated 
version of the near field results with
\be{0} 
|X_1- X_2|_{\infty} =  \frac{v_{\tbox{E}}}{\sqrt{g^{(n)}}}
\ee

\section{The expressions for $G$}

We are now ready to calculate $G$ from Eq.(\ref{e15}).
There are of course $G^{31}$ and $G^{32}$ but 
the expressions look alike so we concentrate, 
say, on the case $X=X_1$, and suppress the 
node indication subscript whenever possible. 
By definition $\mathcal{F}
=-\partial \mathcal{H}/\partial X
=\delta(\hat{x}-x_1)$. 
The current operator has been defined 
in Eq.(\ref{e2}), but we still have the freedom 
to set $x_0$ as we want. So the natural 
choice for sake of calculation, 
is obviously $x=x_1$. We shall expand later 
on the results that would be obtained 
if the current were measured via a different 
section. The matrix elements of the operators 
involved are 
\be{0}
\label{F1Definition}
\mathcal{F}_{nm} 
&=& 
-\frac{2}{L} 
\sqrt{q_{\tbox{D}}^{(n)} q_{\tbox{D}}^{(m)} }
\ \sin(\varphi_{\tbox{D}}^{(n)}) 
\ \sin(\varphi_{\tbox{D}}^{(m)}) 
\\
\nonumber
\mathcal{I}_{nm} 
&=& 
i\frac{e}{\mass L} 
\sqrt{ q_{\tbox{D}}^{(n)} q_{\tbox{D}}^{(m)} } 
\left[
\frac{k_n{+}k_m}{2} \sin{(\varphi_{\tbox{D}}^{(m)} {-} \varphi_{\tbox{D}}^{(n)})}  
+ 
\frac{k_n{-}k_m}{2} \sin{(\varphi_{\tbox{D}}^{(m)} {+} \varphi_{\tbox{D}}^{(n)})}
\right]
\ee
One should notice that $\mathcal{F}_{nm}$ is real and symmetric 
with respect to $n\leftrightarrow m$ interchange, 
while $\mathcal{I}_{nm}$ is antisymmetric and purely 
imaginary as implied by time reversal symmetry.
Once we sum  Eq.(\ref{e15}) over all the occupied levels, 
$nm$ terms cancel with $mn$ terms.  
Within the framework of the ``two level approximation" 
the only remaining term involves the occupied level $n$ 
and the next empty level $m=n+1$
\be{0}
\label{G1}
G^{31}(X_1, X_2) =  2 \frac{ \mathcal{F}_{mn} \ \im[\mathcal{I}_{nm}] } { \Delta_s^2 }
\ee
We recall the following expressions:  
\be{0}
\sqrt{q_{\tbox{D}}^{(+)} q_{\tbox{D}}^{(-)}} 
\ \ &\approx& \ \ \frac{b}{2}|\sin(\theta)|
\\  
\varphi_{\tbox{D}}^{(+)} 
\ \ \approx \ \ \varphi_{\tbox{D}}^{(-)} 
\ \ &\approx& \ \ \left[\frac{1}{4}g\right]^{1/2} 
\\
|\varphi_{\tbox{D}}^{(+)} - \varphi_{\tbox{D}}^{(-)}| 
\ \ &\approx& \ \ 
\frac{2}{\sqrt{b}} \left[\frac{1}{4}g\right]^{1/2}
\frac{1}{|\sin(\theta)|}
\ee
Upon substitution we realize that both in the near 
and in the far field we can neglect the second 
term in the expression for $\mathcal{I}_{nm}$. Consequently  
\be{0}
G^{31}(X_1, X_2) 
\ = \
-\frac{1}{4}\left(gb\right)^{3/2}
\frac{ev_{\tbox{E}}}{L^2} 
\ \frac{1}{\Delta_s^2}
\ |\sin(\theta)| 
\ = \
-\frac{2 \ ev_{\tbox{E}}}{b\sqrt{g}} \ 
\frac{|X_1 - X_2|}{||\bm{X}-\bm{X}^{(r)}||^3}
\ee
We observe that in the near field, 
where $||\bm{X}-\bm{X}^{(r)}||$ is essentially 
the Euclidean measure of distance, 
we get the field of a Dirac monopole 
as expected. But as we go to the far field 
the $|X_1-X_2|$ contribution saturates 
as explained in the previous section.

\section{The calculation of $Q$}

Having found $G^{31}$ and a similar 
expression for $G^{32}$ we can perform 
the integration of Eq.(\ref{e3}) in order 
to obtain $Q$. We already pointed out 
that most of the contribution 
in the regions of our interest 
comes from the avoided crossings. 
In the near field  it is most convenient 
to make the integration along 
$X_1-X_2=\pm\const$ segments,  
while in case of the far field 
we make the integration along   
$X_2=\infty$ and $X_1=\infty$ segments.

It is not difficult to realize that 
in the near field calculation both 
segments ($X_1-X_2=\pm\const$) give 
the same contribution. This means that 
we simply have to do one of the integral 
and to multiply by~$2$. 
But in the far field one should 
be more careful. If we measure 
the current in the {\em middle} of the dot 
(as indeed done in the numerics) 
then the same rule applies. But if 
we measure the current (say) at node~$1$, 
then the predominant contribution to $Q$    
comes obviously from the $X_2=\infty$ segment, 
so the result should not be multiplied by~$2$. 
Performing a straightforward calculation 
of the $dX$ integral, using   
\be{0}
\int_{-\infty}^{+\infty}
\frac{dx}{\left( x^2 + a^2 \right)^{3/2}} = \frac{2}{a^2}
\ee
and taking the above discussion into account, 
we get in the near field 
\be{81}
Q \ \ \approx \ \ \frac{X^{(r)}}{|X_1-X_2|} e
\ee
One can show that this result is in agreement 
with the rough estimate of Eq.(\ref{e23}). 
However, in the far field we have to substitute 
the saturated value of  $|X_1-X_2|$ 
leading to the result $Q \approx e$.

The calculation in the far field does not care 
whether the pumping cycle encircle a $\Phi=0$ 
degeneracy or a  $\Phi=\pi$ degeneracy. 
It is only in the near field of $X^{(r)}$ 
that we see the difference. This is clearly 
confirmed by the numerics (Fig.4). 
But a closer look reveals that the far field 
numerical result for $Q$ is somewhat 
smaller than~$1$. This might look like 
a contradiction with respect to the general expectations. 
The resolution of this puzzle is related 
to the limitation of the far field perturbative 
treatment. Within the framework of this 
treatment $\Theta^{(n)}$ changes 
from $\Theta^{(n)}=\pi$ to $\Theta^{(n)}=0$ 
as we lower (say) the $X_1$ barrier.
But in fact we know from section~8 that $\Theta$ 
is bounded. This means that not all the 
the probability gets into the dot region. 
Consequently,  if we integrate 
along the $X_2=\infty$ segment, 
we expect to get $Q<e$ as observed. 
On the other hand, in case of a full 
pumping cycle, we have to cross 
from the $X_2=\infty$ segment  
to the  $X_1=\infty$ segment. 
This was neglected in our calculation. 
Thus if we had a full cycle we would 
expect to get $Q\approx e$ in the far field 
as implied by the Dirac chains picture.

\section{Discussion}

We were able to derive an estimate for $Q$ 
in the case where the pumping cycle 
is dominated by a single degeneracy. 
Within this framework 
we can still distinguish between 
near field (where $Q\gg e$) 
and far field (where $Q\sim e$) regions.   
Our results are in agreement with 
those of Ref.\cite{pmc}. We note that 
an optional derivation of the near field 
limit has been introduced in Ref.\cite{pMB} 
using an extension of the BPT scattering 
formalism. But the latter derivation  
was not suitable to reproduce the far field  
result because it has been {\em assumed} there  
that the charge cannot accumulate in the dot region.

It should be re-emphasized that the results 
that we have obtained assume that the pumping 
cycle is dominated by a single degeneracy. 
In a follow up work \cite{pmm} we shall 
discuss the case where the charge transport 
involves many levels, such that the 
contribution of neighboring levels (in Eq.(\ref{e15})) 
is negligible compared with contribution 
that comes from $|m-n|>1$ levels.

The results that we obtain for a closed geometry are very 
different from those that are obtained for an open 
geometry. This is because the motion of the electron 
is ``recycled". In a more technical language this means 
that multiple rounds should be taken into account in 
the calculation of correlation functions. 
Refs.\cite{pmo,pme} further discuss how the Kubo formula 
can be used in order to interpolate these two extreme 
circumstances.

It should be clear that adiabatic transport becomes  
counter-intuitive if one adopts a misleading 
zero order point of view of the adiabatic process.
Moreover, even within the ``two level approximation" 
it would not be correct to say that $Q$ is determined 
by {\em peristaltic} mechanism: 
namely, it is not correct to say 
that charge transfer is regulated   
by the Landau-Zener transitions.

A {\em peristaltic} mechanism would imply $Q\sim e$.  
In the near field we have  $Q\gg e$ so we do not have 
such  mechanism for sure. This is also reflected 
by having the same~$I$ at both nodes as discussed 
in the paragraph of Eq.(\ref{e81}).
However, even in the far field, 
where the peristaltic picture seems natural, 
we have realized that it is an over simplification:    
also in the case of a far field cycle a finite 
fraction of $Q$ is contributed during 
the intermediate stages of the pumping cycle.

\appendix
\section{Finding the degeneracy points}

If system is symmetric ($X_1=X_2=X$) then we can distinguish between 
odd and even states leading to the following eigenvalue equations 
\be{101}
\cot(kL_{\tbox{D}}/2) + \cot(kL_{\tbox{W}}/2) 
&= -\frac{2\mass}{k}X   
& \,\,\, \mbox{odd states} 
\\ \label{e102} 
\tan(kL_{\tbox{D}}/2) + \tan(kL_{\tbox{W}}/2) 
&= +\frac{2\mass}{k}X   
& \,\,\, \mbox{even states}
\ee
As we lower $X$ we have an exact crossing whenever 
a dot state crosses a wire state with the opposite parity. 
We have an avoided crossing whenever a dot state 
tries to cross a wire state with the same parity.  
The later becomes an exact crossing  
if the flux through the ring is half integer.

We can determine the degeneracy points by 
equating (\ref{e101}) with (\ref{e102}).  
This gives $\sin(kL_{\tbox{W}})=-\sin(kL_{\tbox{D}})$. 
For half integer flux it is 
convenient to use delta gauge on the middle of the wire, 
so as to get there $\pi$ phase jump boundary conditions. 
This implies that in the above equation we make 
the replacement $(kL/2) \mapsto (kL/2) + (\pi/2)$, 
hence getting the degeneracy condition  
$\sin(kL_{\tbox{W}})=+\sin(kL_{\tbox{D}})$. 
We therefore conclude that we have degeneracies for 
\be{0}
k_r \ \ = \ \ \frac{\pi}{L_{\tbox{W}}-L_{\tbox{D}}} n_r
\ee 
They are categorized into $\Phi{=}0$ degeneracies 
for $n_r=1,3,5,\dots$ and $\Phi{=}\pi$ degeneracies 
for $n_r=2,4,6,\dots$. 
Their location is $(X^{(r)},X^{(r)})$ where   
\be{0}
X^{(r)} \ \ = \ \ -\frac{k_r}{\mass}\cot(k_rL_{\tbox{D}}) 
\ee
Accordingly
\be{0}
g^{(r)} \ \ = \ \ g(X^{(r)})
\ \ = \ \ \sin^2(k_rL_{\tbox{D}}) 
\ \ = \ \  \sin^2(k_rL_{\tbox{W}})
\ee
At a degeneracy point the mixing parameter $\Theta$ 
that characterizes the odd and the even states 
attains the extremal values which are allowed by Eq.(\ref{e35}). 
This can be verified by deducing $q_{\tbox{W}}/q_{\tbox{D}}$ 
from Eq.(\ref{e31}) with ${\varphi_{\tbox{D}}=k_rL_{\tbox{D}}/2}$ for the 
odd state and ${\varphi_{\tbox{D}}=(\pi/2)+k_rL_{\tbox{D}}/2}$ 
for the even state.

\ack

D.C. thanks M.~Moskalets and M.~B{\"u}ttiker 
for discussions that had motivated this work,  
and Y.~Oreg for urging clarification of the 
formal result. We are grateful to 
T.~Kottos, H.~Schanz and G.~Rosenberg 
for helpful suggestions and help with the 
numerical procedure. The research was supported 
by the Israel Science Foundation (grant No.11/02),
and by a grant from the GIF, the German-Israeli Foundation 
for Scientific Research and Development.


\newpage

\mpg{


\Cn{\putgraph[width=0.9\hsize]{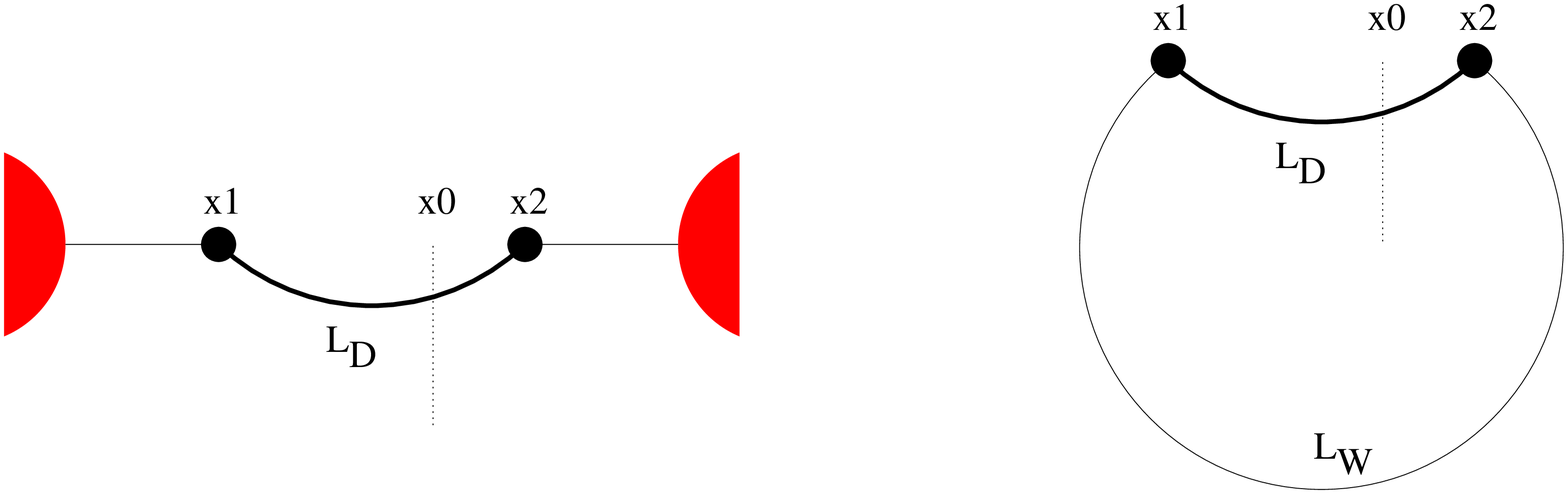}} 

{\footnotesize {\bf FIG. 1.} 
Illustration of the model system. 
The two barrier pumping device 
is used in two different configurations. 
Left panel: open lead geometry;  
Right panel: closed ring geometry.
The barriers are located 
at the nodes $x_1$ and $x_2$ 
while the current is measured 
through the section at $x_0$. }

}

\ \\ \ \\

\mpg{


\Cn{
\putgraph[height=0.3\hsize]{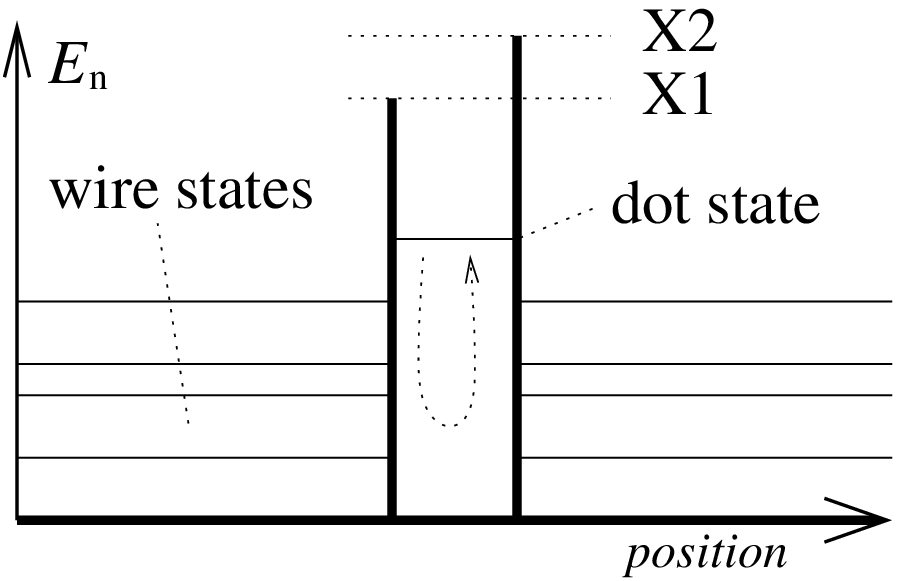}
\putgraph[height=0.3\hsize]{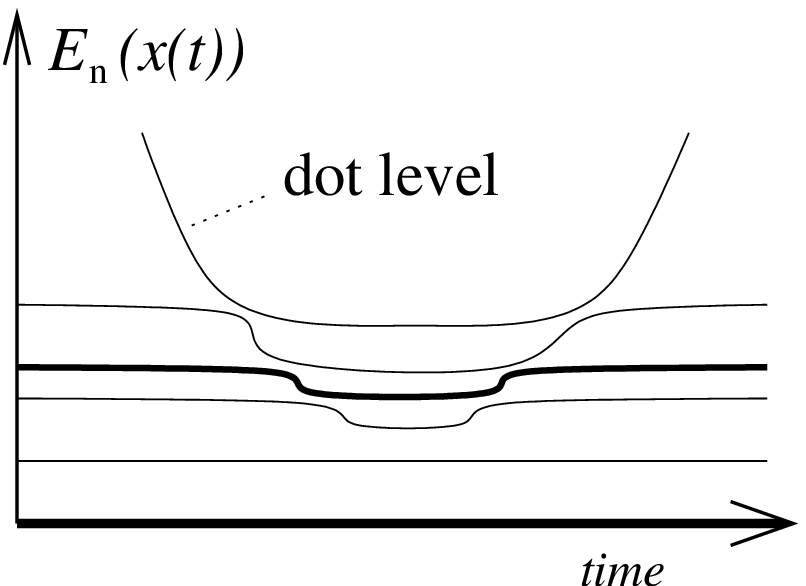} 
}

\Cn{
\putgraph[width=0.92\hsize]{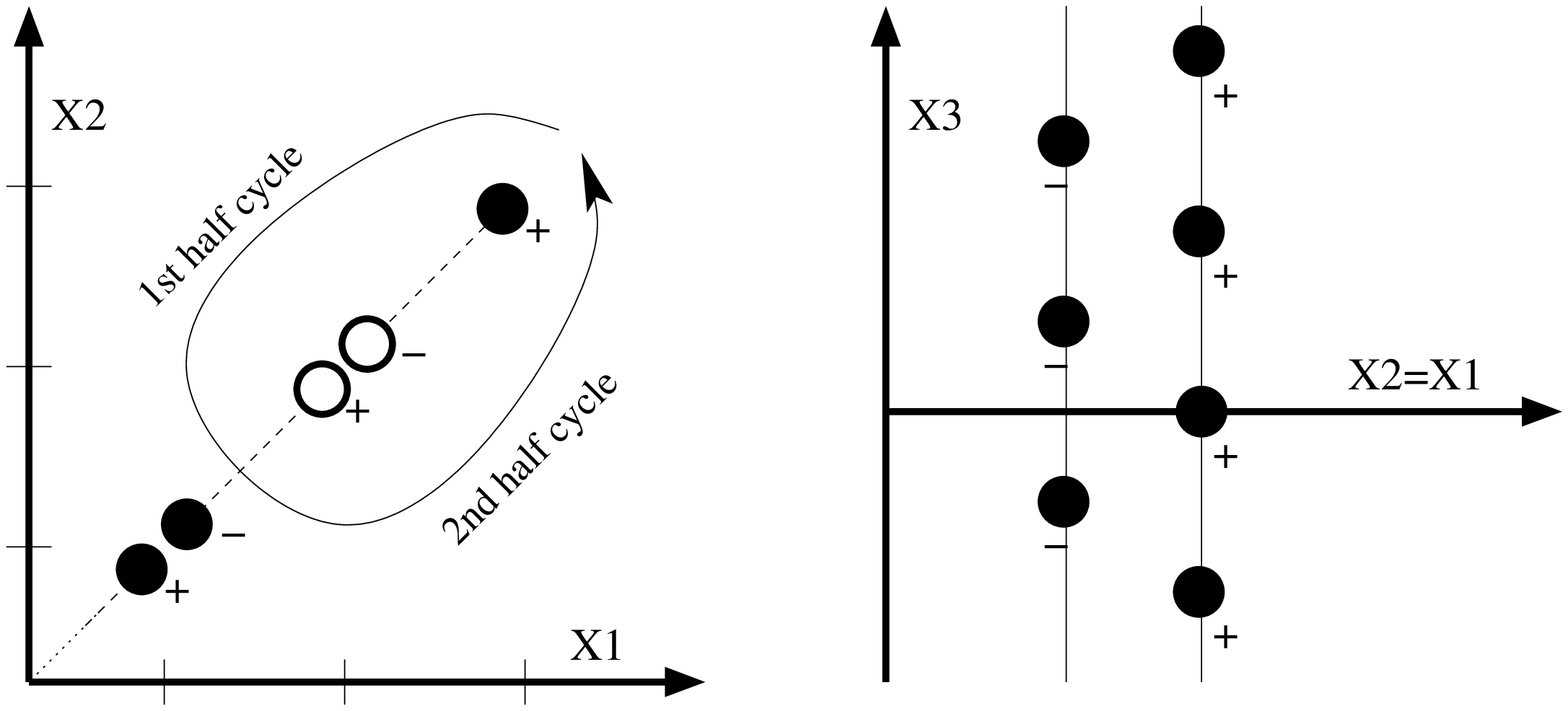}
}

{\footnotesize {\bf FIG 2.}
(a) Upper left: The energy levels of a ring 
with two barriers, at the beginning of the pumping cycle. 
It is assumed that the three lower levels are occupied. 
(b) Upper right: The adiabatic levels as a function 
of time during the pumping cycle. 
(c) Lower Left: The $(X_1,X_2)$ locations of 
the Dirac chains of the $3$ occupied levels. 
Filled (hollow) circles imply that there 
is (no) monopole in the pumping plane. 
Note that for sake of illustration overlapping 
chains are displaced from each other.
The pumping cycle encircles $2+1$ Dirac chains 
that are associated with the 3rd and 2nd levels respectively.   
(d) Lower right: The $2$ Dirac chains 
that are associated with the 3rd level.}

}

\ \\

\mpg{


\putgraph[width=0.45\hsize]{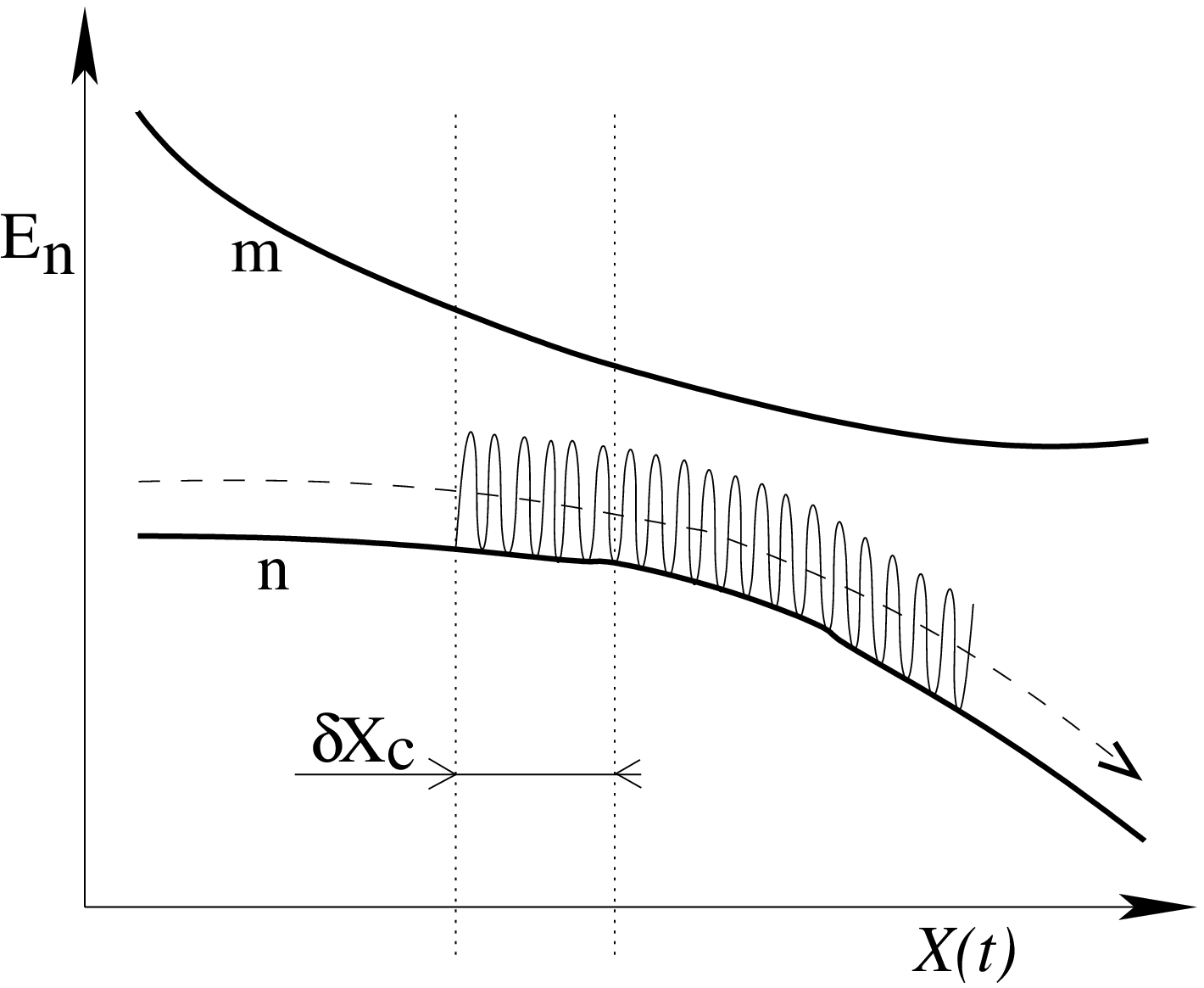} 
\mpg[0.5\hsize]{
{\footnotesize {\bf FIG. 3.} }
Cartoon of the adiabatic evolution 
within the framework of the two level approximation. 
The system is prepared in level $n$, 
and the nearby empty level is ${m=n+1}$. 
The control parameter $X(t)$ is being changed 
slowly, and therefore the system ``oscillates" 
around the first order adiabatic solution. 
The energy of the latter is illustrated 
by a dashed line. Note that the identity 
of the adiabatic state changes gradually, 
and can be regarded as constant only  
on scales $\delta X \ll \delta X_c$.}  
   
}

\ \\ \ \\

\mpg{


\Cn{
\mpg[0.4cm]{(a) \\ \vspace*{4cm} }
\putgraph[width=0.5\hsize]{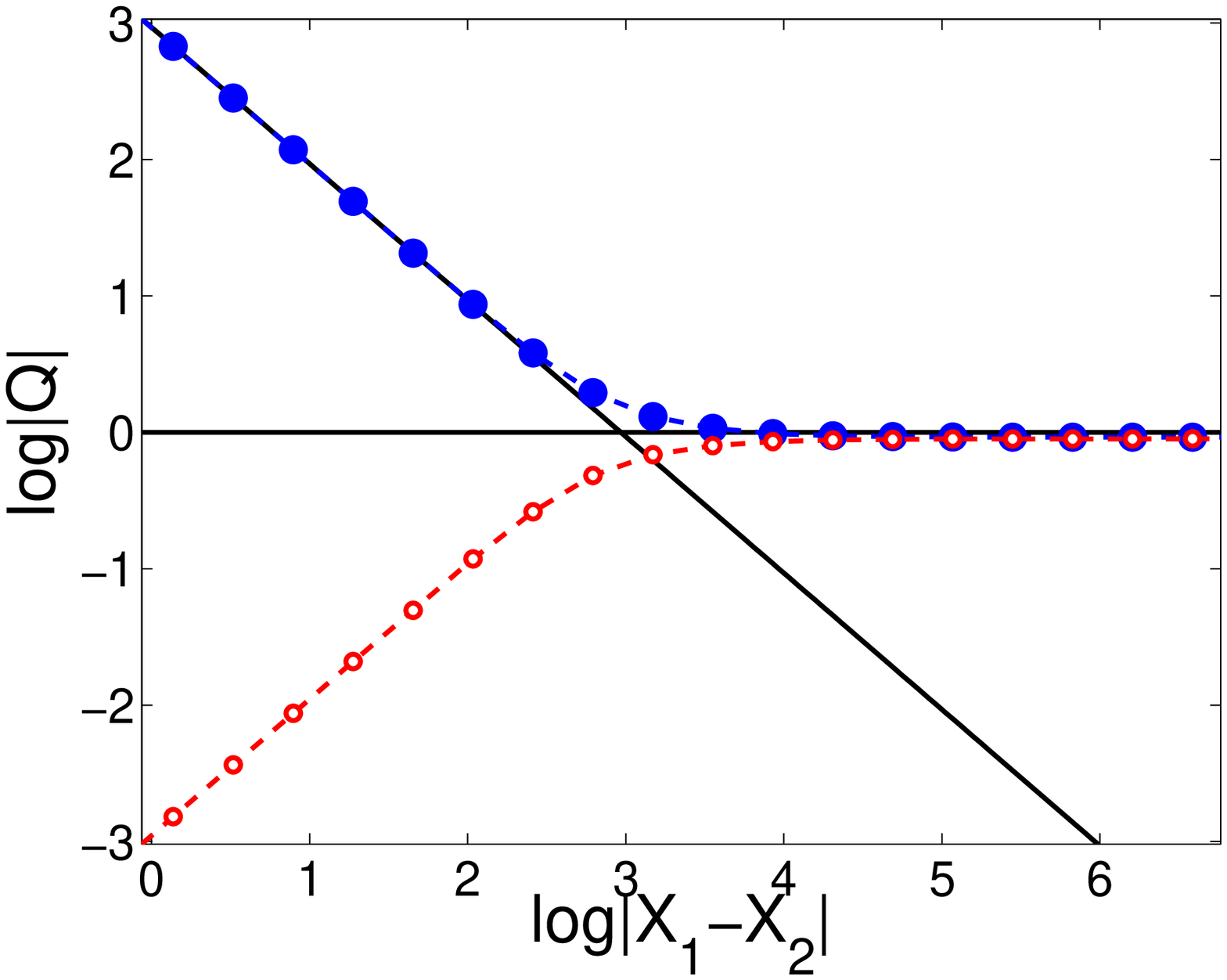}
}

\Cn{
\mpg[0.4cm]{(b) \\ \vspace*{2cm} }
\putgraph[height=0.25\hsize]{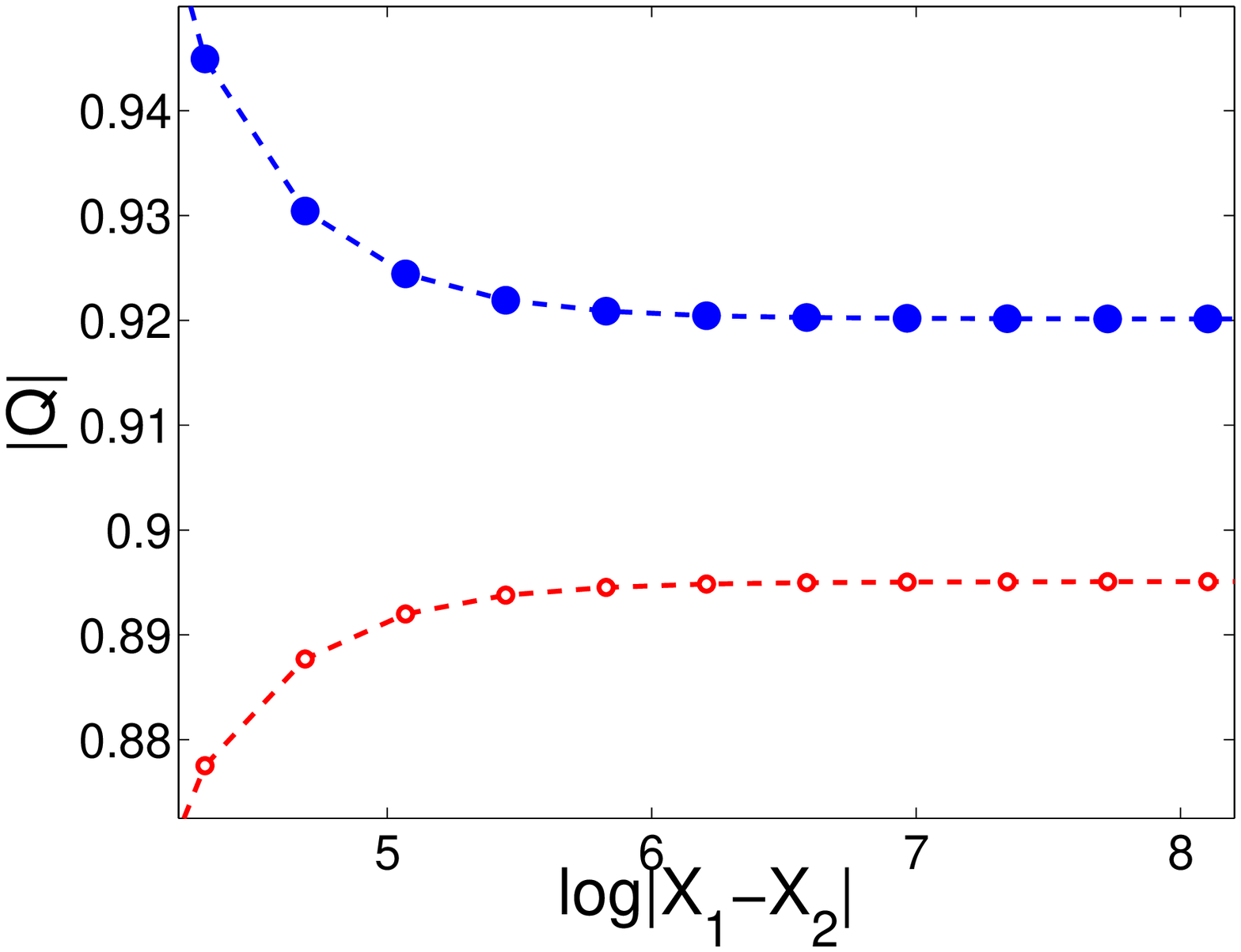}
\hspace*{1cm}
\mpg[0.4cm]{(c) \\ \vspace*{2cm} }
\putgraph[height=0.25\hsize]{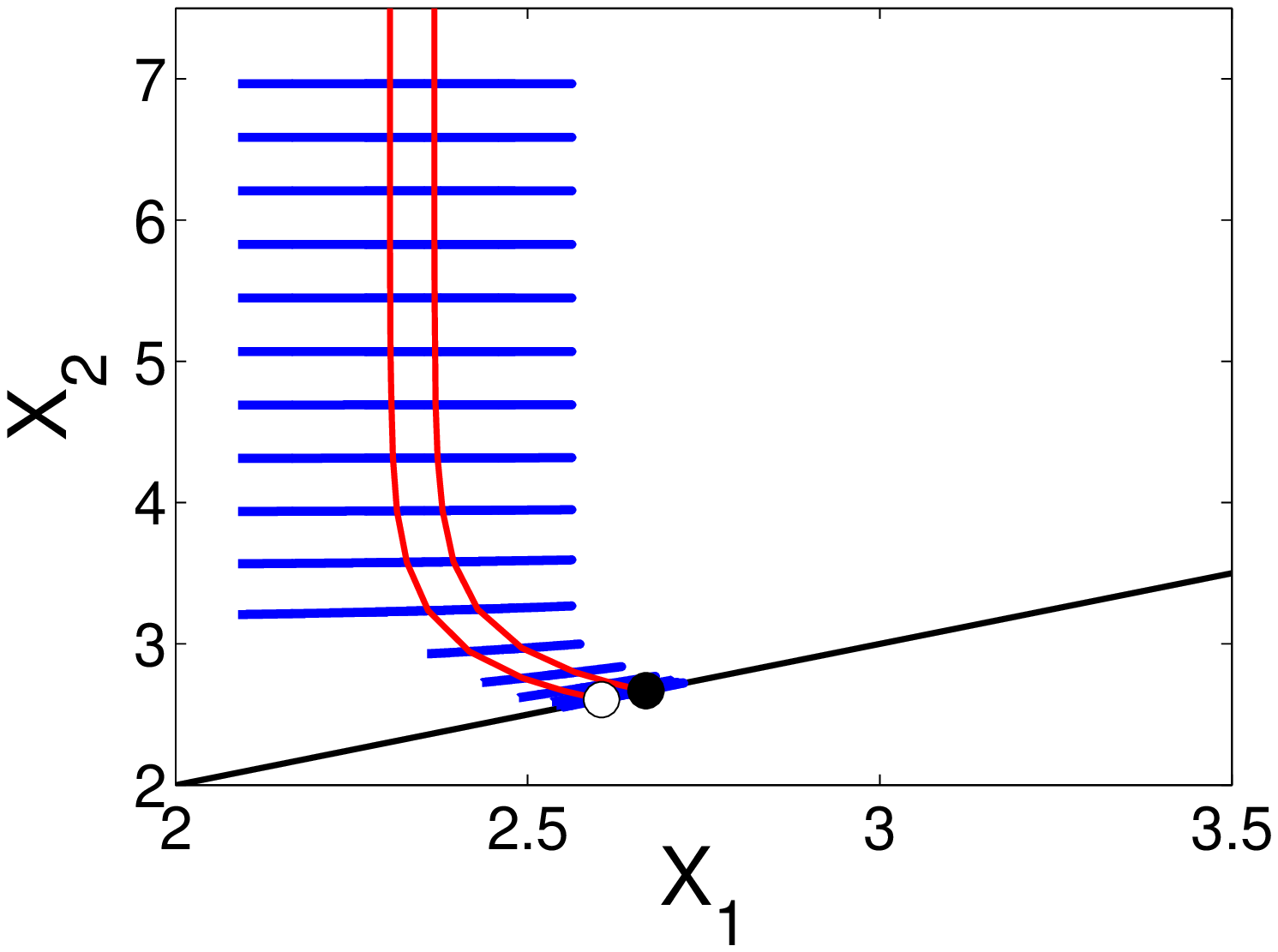}
}

{\footnotesize {\bf FIG. 4.} }
Panels (a) and (b): 
Numerical calculation of the pumped charge $Q$.
The model parameters are ${L_{\tbox{D}}=\mass=e=1}$   
and ${L_{\tbox{W}} = 3000.43}$. The current $I$ 
is measured at the middle of the dot. 
The numerical integration is carried out along 
the segments which are indicated in panel (c),   
and the results are multiplied by~2 so as to include 
the equal contribution that comes from the second 
half of the cycle.  There are two sets of data points.
One set (filled circles) is for pumping cycles that encircle 
an in-plane degeneracy point ($n_r=2993$).  
A second set (hollow circles) is for pumping cycles 
that encircle an off-plane degeneracy point ($n_r=2992$).
The location of the avoided crossing for each 
data set is indicated by the solid lines in panel (c).
The near and that far field approximations 
that we derive for~$Q$ are indicated by the 
the solid lines in panel~(a). The zoom in panel~(b) 
reveals that~$Q$ in the far field is in fact slightly 
less then~$1$, which is explained in section~13.
}

\mpg{


\Cn{
\mpg[0.4cm]{(a) \\ \vspace*{5cm} }
\putgraph[height=0.6\hsize]{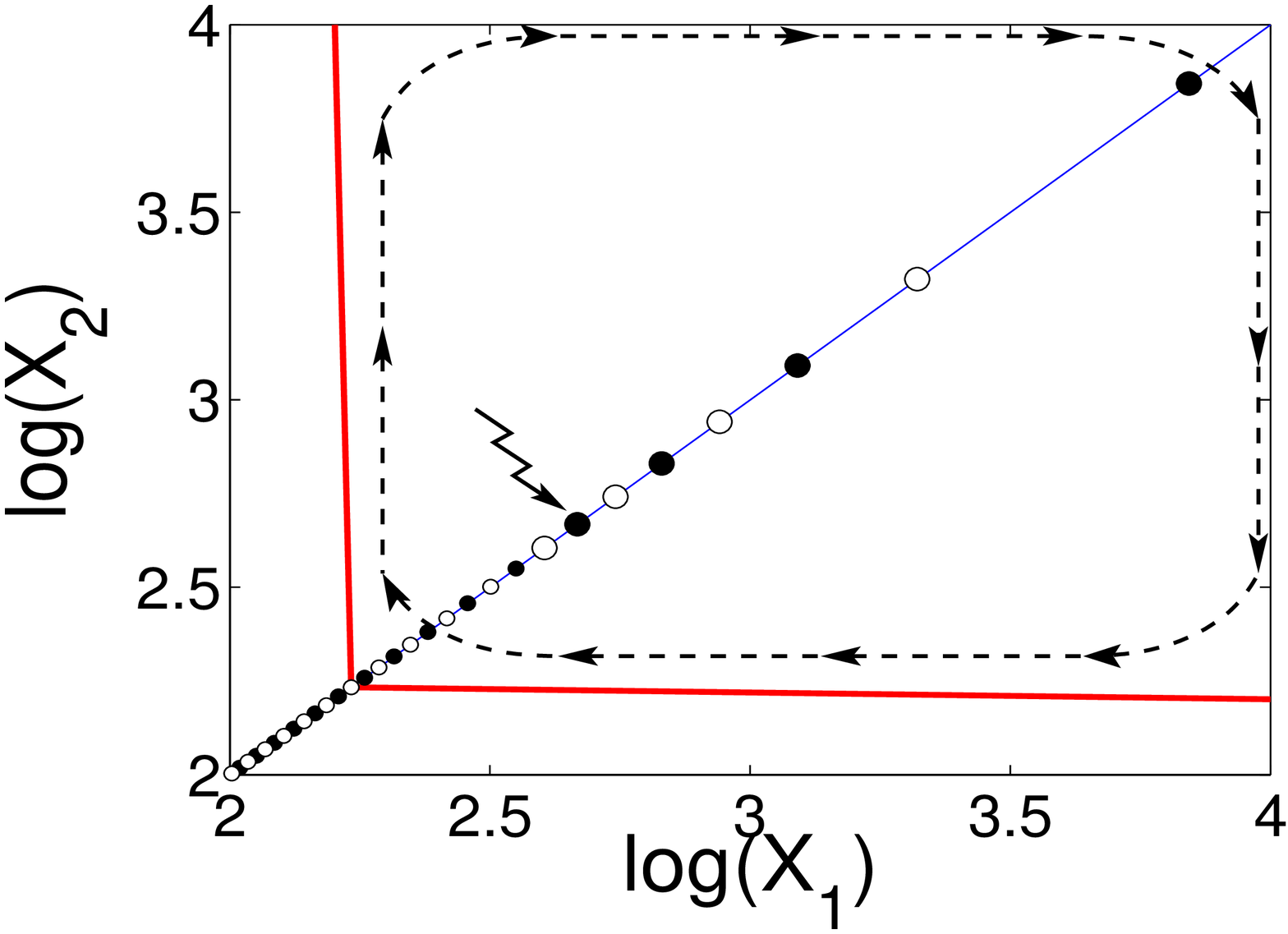}
}

\Cn{
\mpg[0.4cm]{(b) \\ \vspace*{3cm} }
\putgraph[height=0.35\hsize]{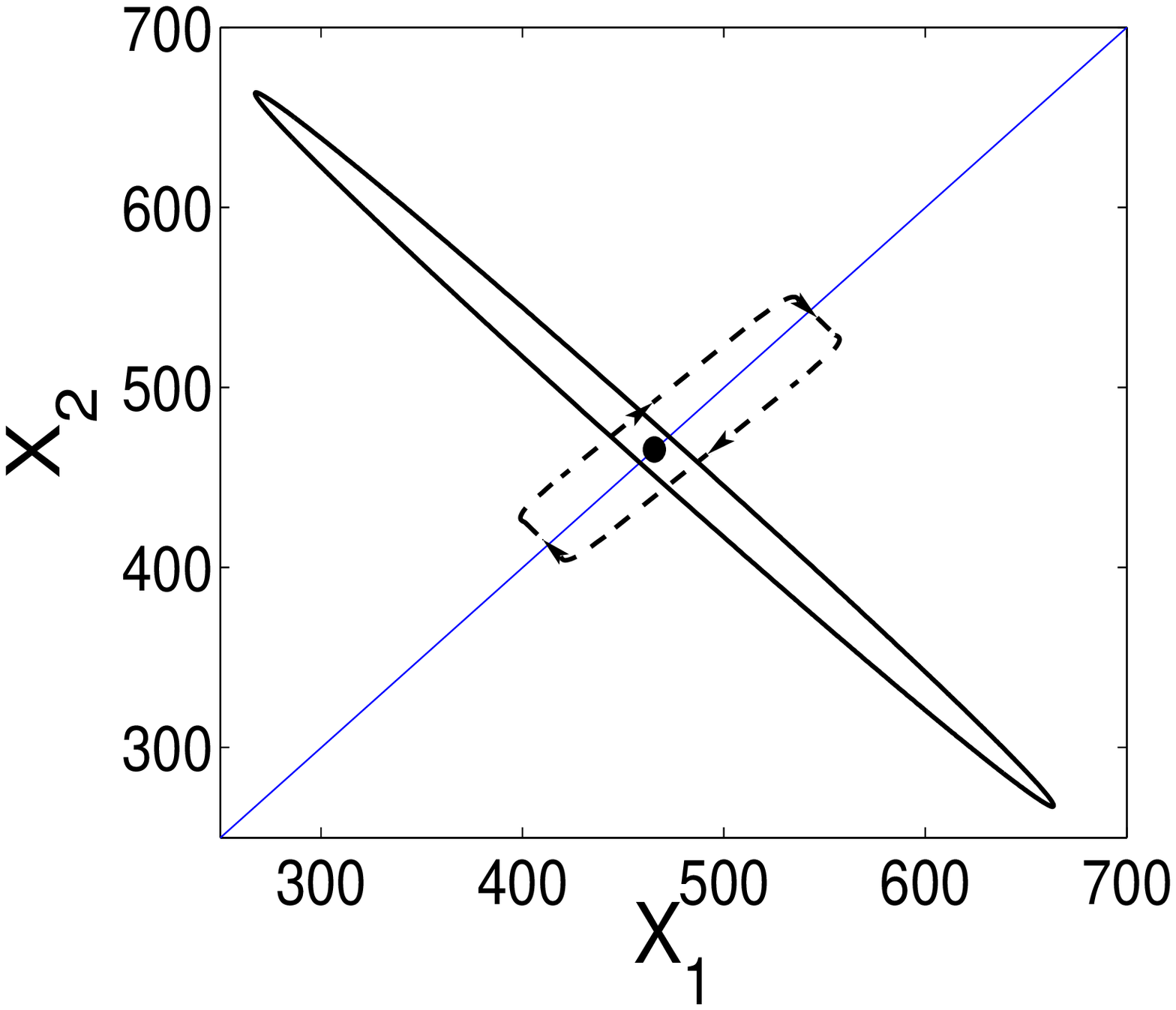}
\hspace*{1cm}
\mpg[0.4cm]{(c) \\ \vspace*{3cm} }
\putgraph[height=0.33\hsize]{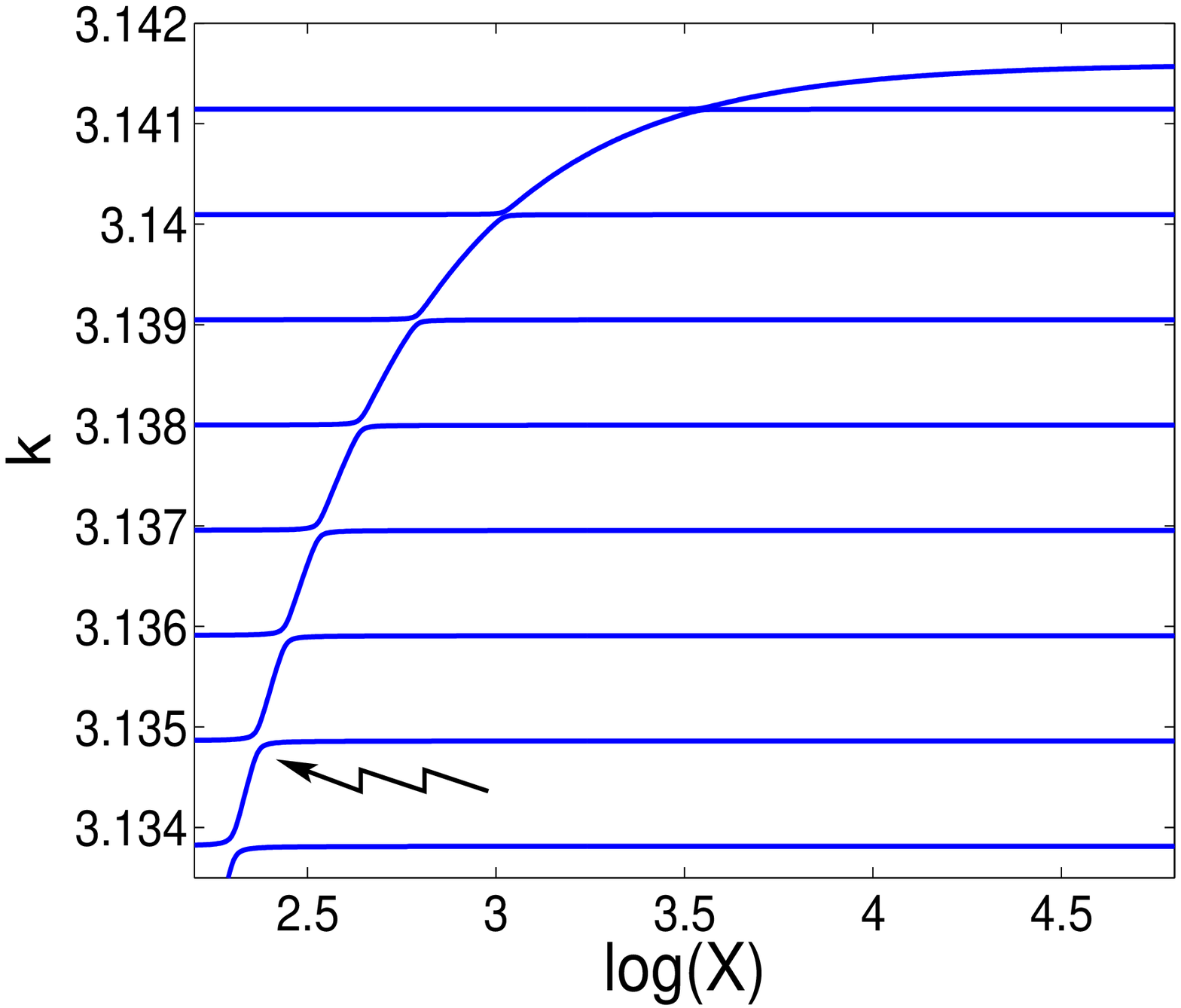}
}

\Cn{
\mpg[0.4cm]{(d) \\ \vspace*{3cm} }
\putgraph[width=0.4\hsize]{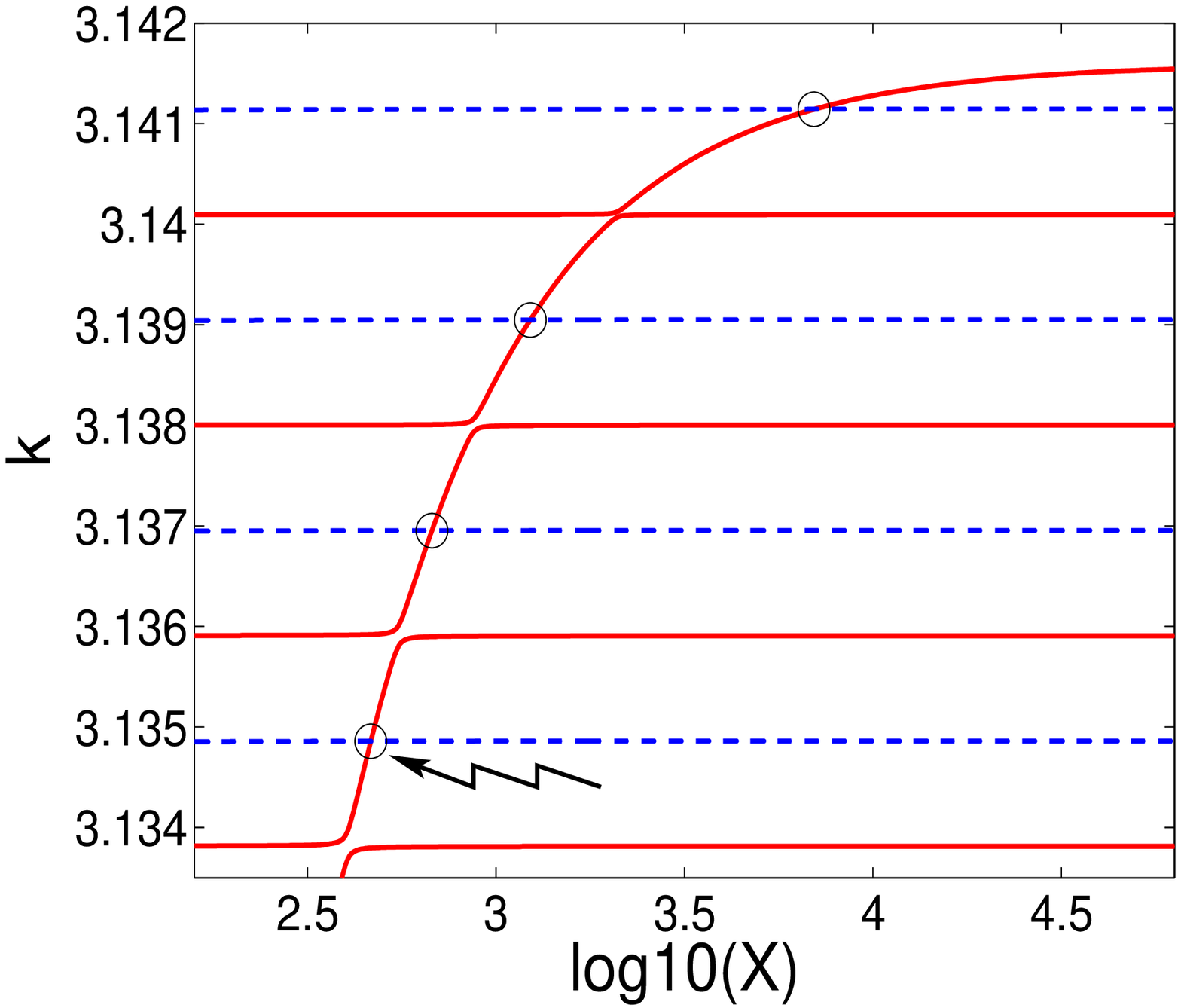}
\hspace*{1cm}
\mpg[0.4cm]{(e) \\ \vspace*{3cm} }
\putgraph[width=0.4\hsize]{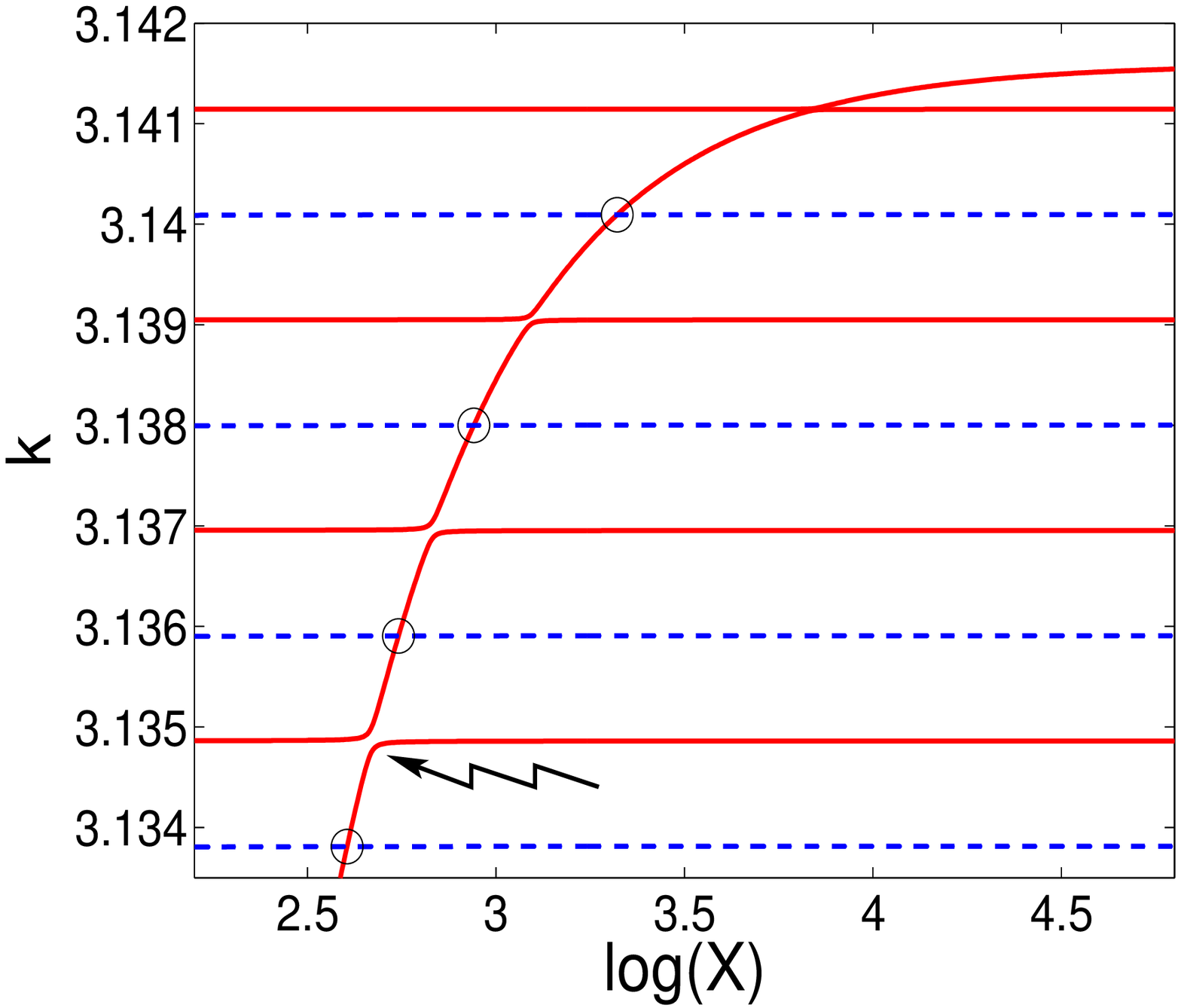}
}

{\footnotesize {\bf FIG. 5.} 
Regions in $\bm{X}$ space. The model parameters 
are the same as in the previous figure. 
Panel (a) displays the region of interest 
as defined in Eq.(\ref{e34}). It is bounded 
by the left and by the bottom solid lines 
which are defined by $g(X) \sim 1/b$.
In-plane and off-plane degeneracy points 
are indicated by filled and hollow circles respectively.  
We indicate by arrow one in-plane degeneracy 
point ($n_r = 2993$). A zoom of its near 
field is displayed in panel (b). The ellipse 
in panel (b) indicates a level splitting 
that equals $\Delta/10$. The dashed lines 
in panels (a) and (b) indicate far field and near 
field pumping cycles respectively.   
In panels (c)-(e) we shows the energy levels ($k_n$) 
along three paths in $\vec{X}$ space, 
which are $(X_1{=}X,X_2{=}\infty, \Phi{=}0)$,  
and $(X_1{=}X,X_2{=}X, \Phi{=}0)$,  
and $(X_1{=}X,X_2{=}X, \Phi{=}\pi)$ respectively.
The degeneracies ${n_r = 2992...2999}$ 
are circled. The arrow indicates the 
representative degeneracy point ${n_r = 2993}$. 
In panels (d) the odd states are indicated 
by dashed lines so as to distinguish 
them from the even states. }

}

\mpg{


\mpg[0.4cm]{(a) \\ \vspace*{4cm} }
\putgraph[width=0.6\hsize]{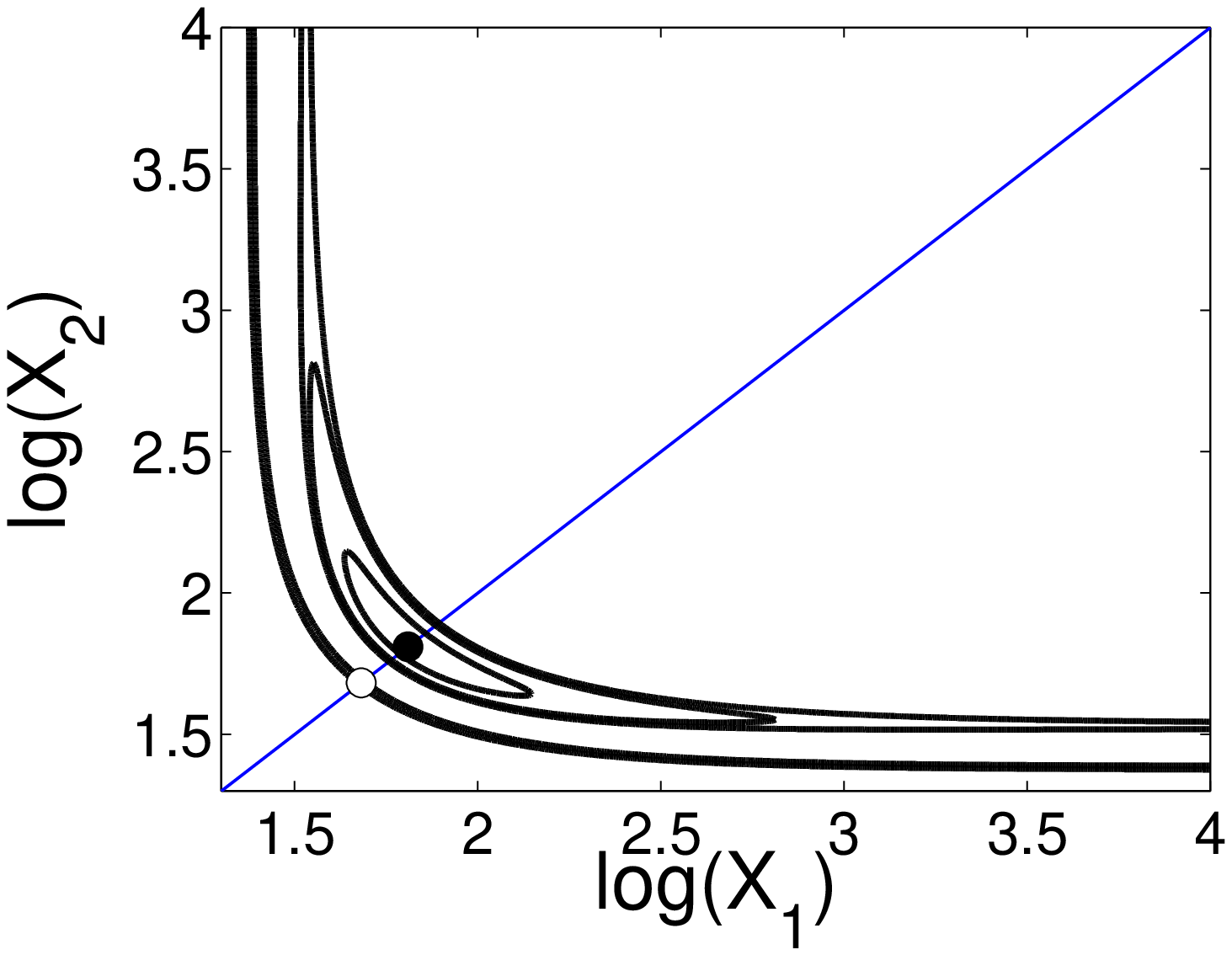} 

\mpg[0.4cm]{(b) \\ \vspace*{4cm} }
\putgraph[width=0.6\hsize]{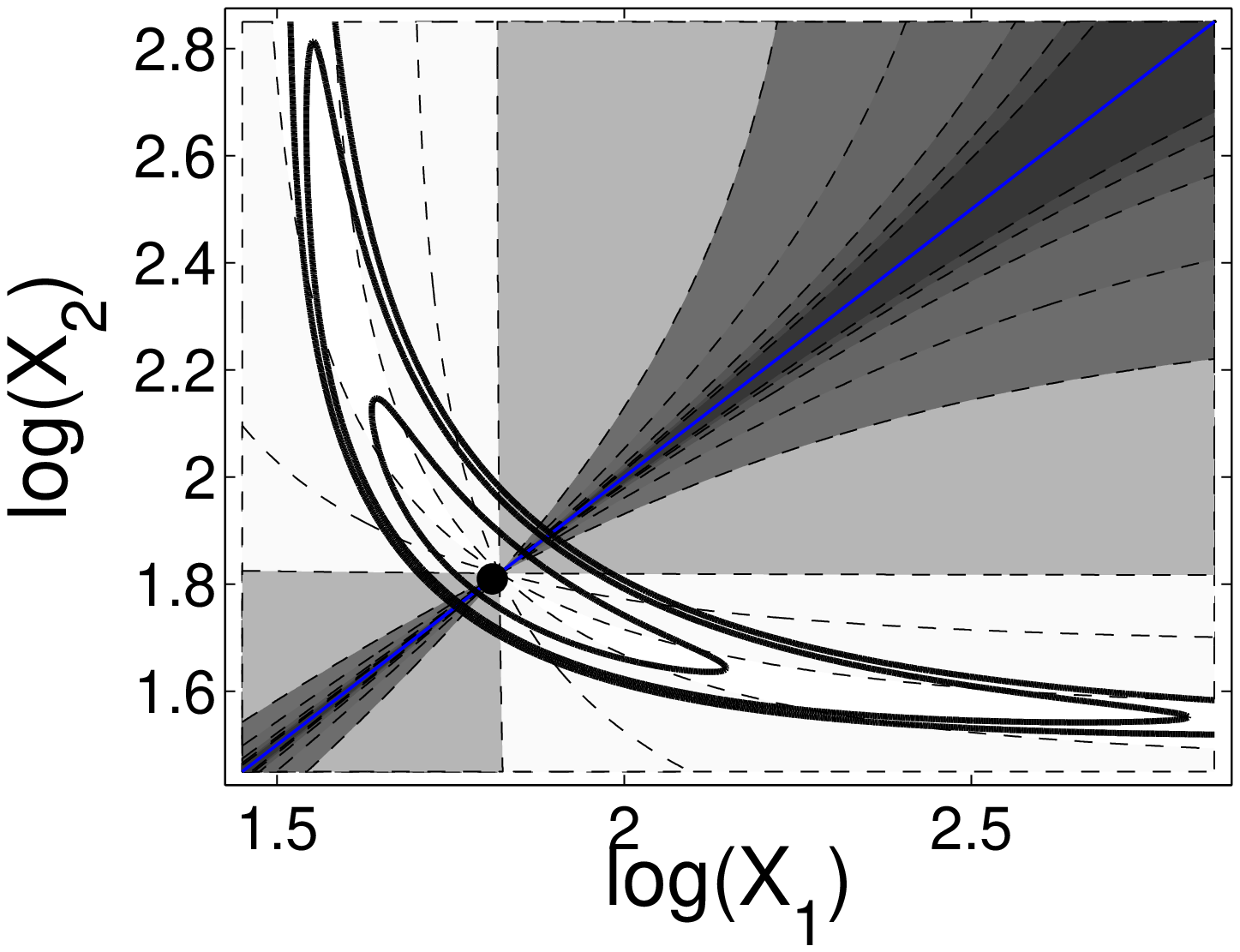}

\mpg[0.4cm]{(c) \\ \vspace*{4cm} }
\putgraph[width=0.6\hsize]{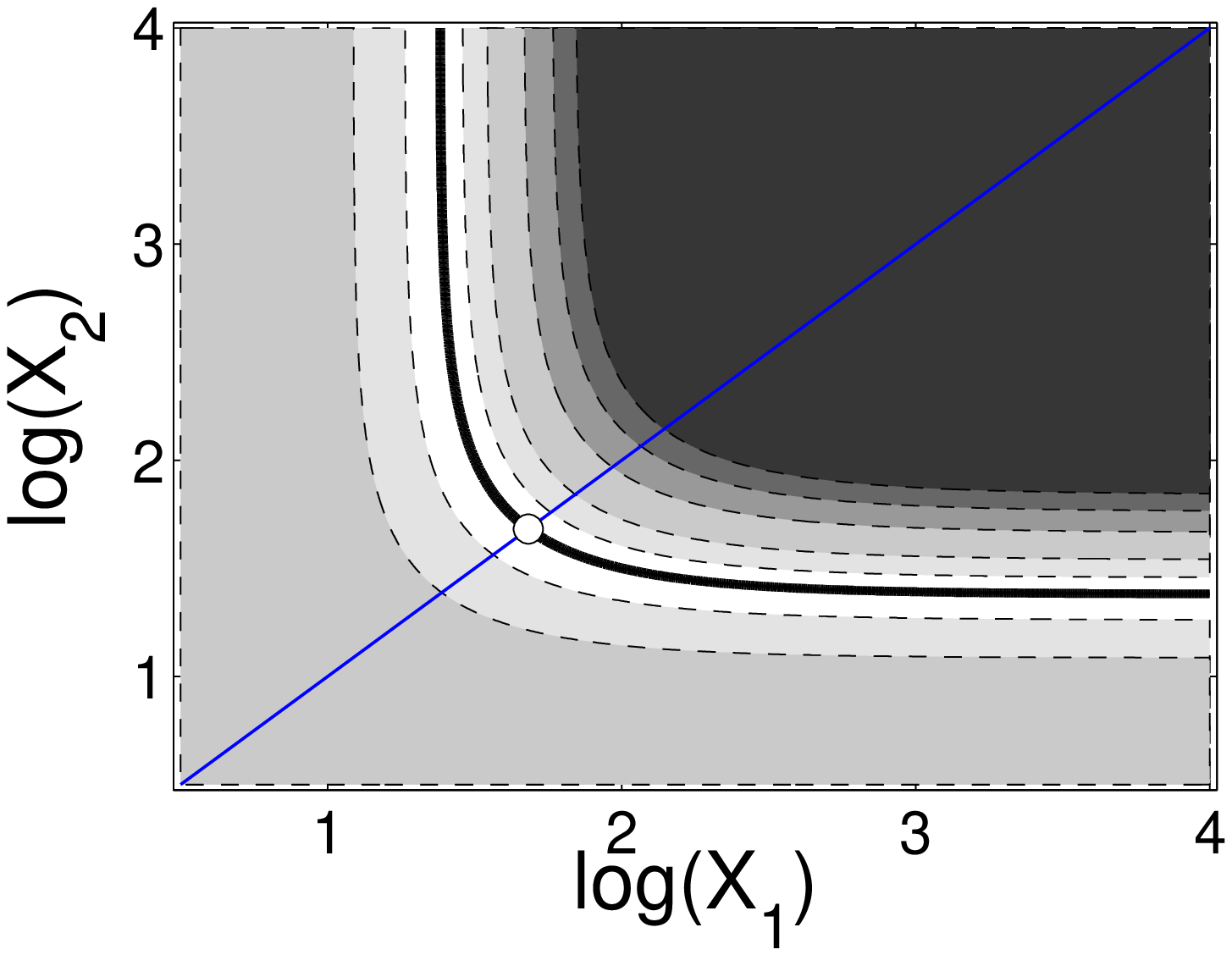}

{\footnotesize {\bf FIG. 6.} 
The energy level splitting and the mixing 
parameter $\Theta$ for two pairs of levels.
The model parameters are ${L_{\tbox{D}}=\mass=e=1}$   
and ${L_{\tbox{W}} = 160.43}$. 
In panel (a) we show the contour lines 
for the energy level splitting of 
the first (even) dot level with an odd wire level ($n = 158$), 
and for the energy level splitting of 
the first (even) dot level with 
an even wire level ($n = 157$).   
The two cases are displayed again 
in panels (b) and (c) respectively 
where we plot both level splitting contours 
(solid lines) and $\Theta$ contours (dashed lines).
In the ``even-odd crossing" case 
we have an in-plane degeneracy, which is 
indicated by a filled circle,  
while the inner most contour line 
is for $\Delta/5$ splitting. 
Note that within the white regions 
the mixing is maximal ($\Theta\sim\pi/2$).   
In the ``even-even avoided crossing" case 
the projection of the off-plane degeneracy point  
is indicated by a hollow circle.} 

%

}

\mpg{


\putgraph[width=0.3\hsize]{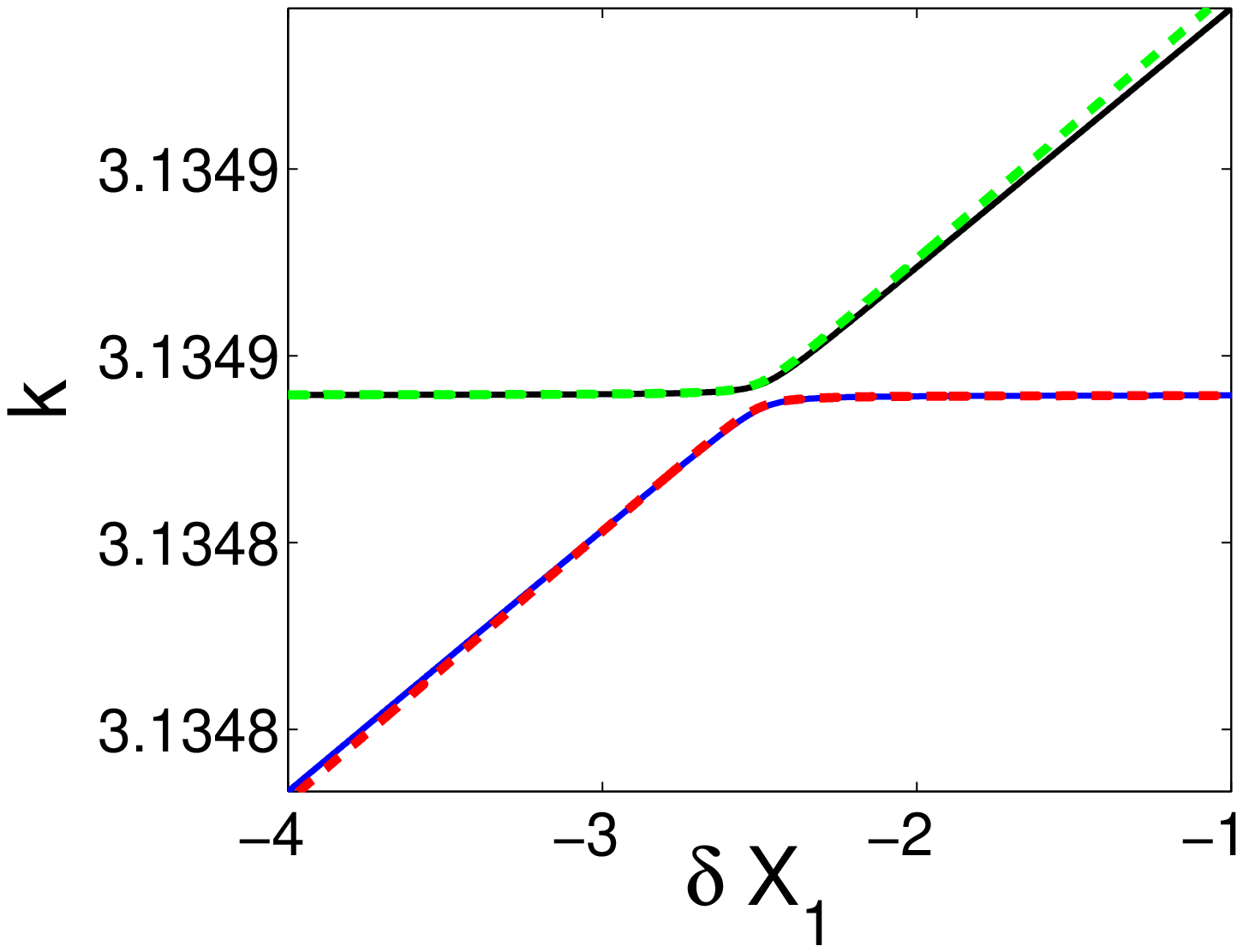}
\putgraph[width=0.3\hsize]{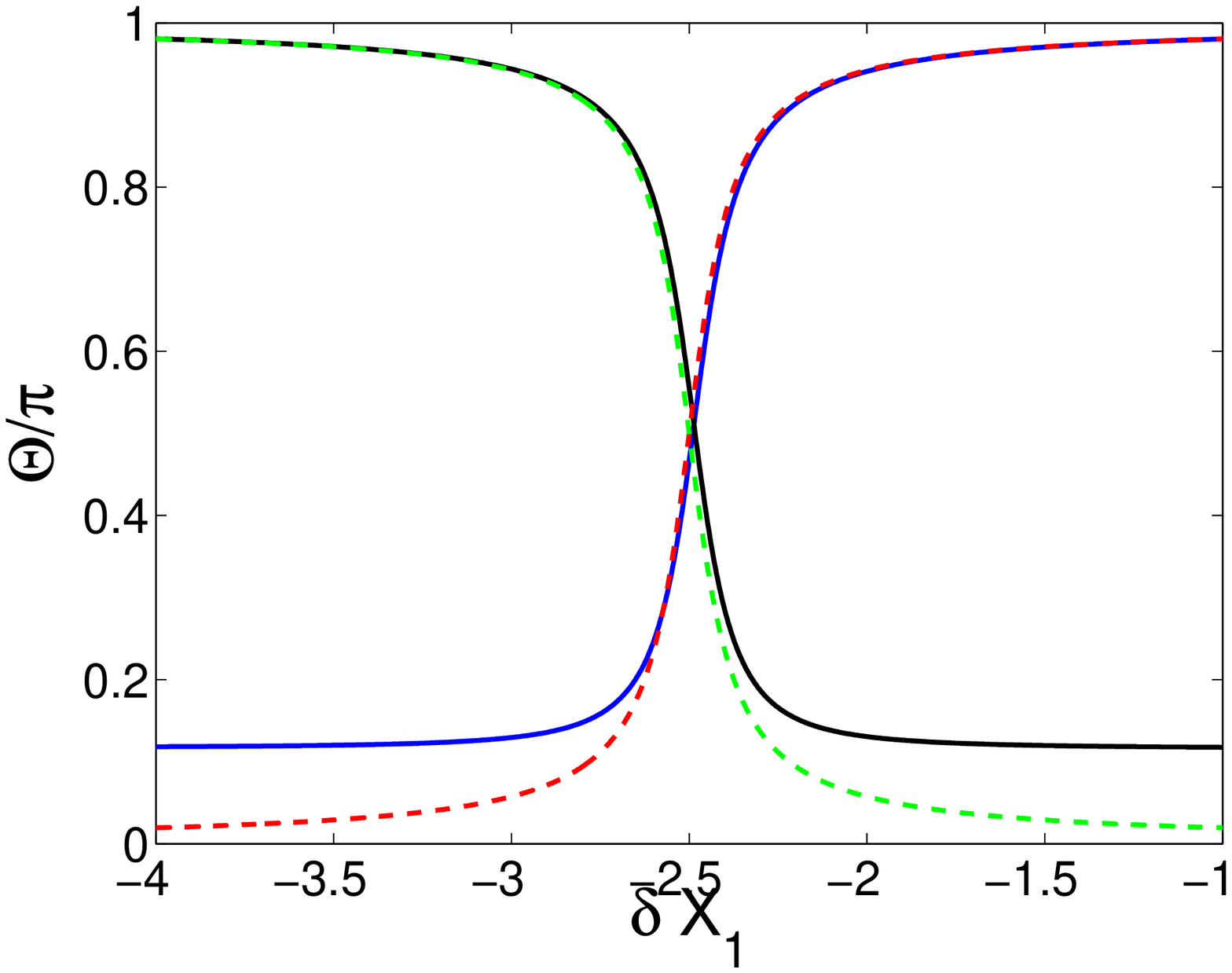}
\putgraph[width=0.3\hsize]{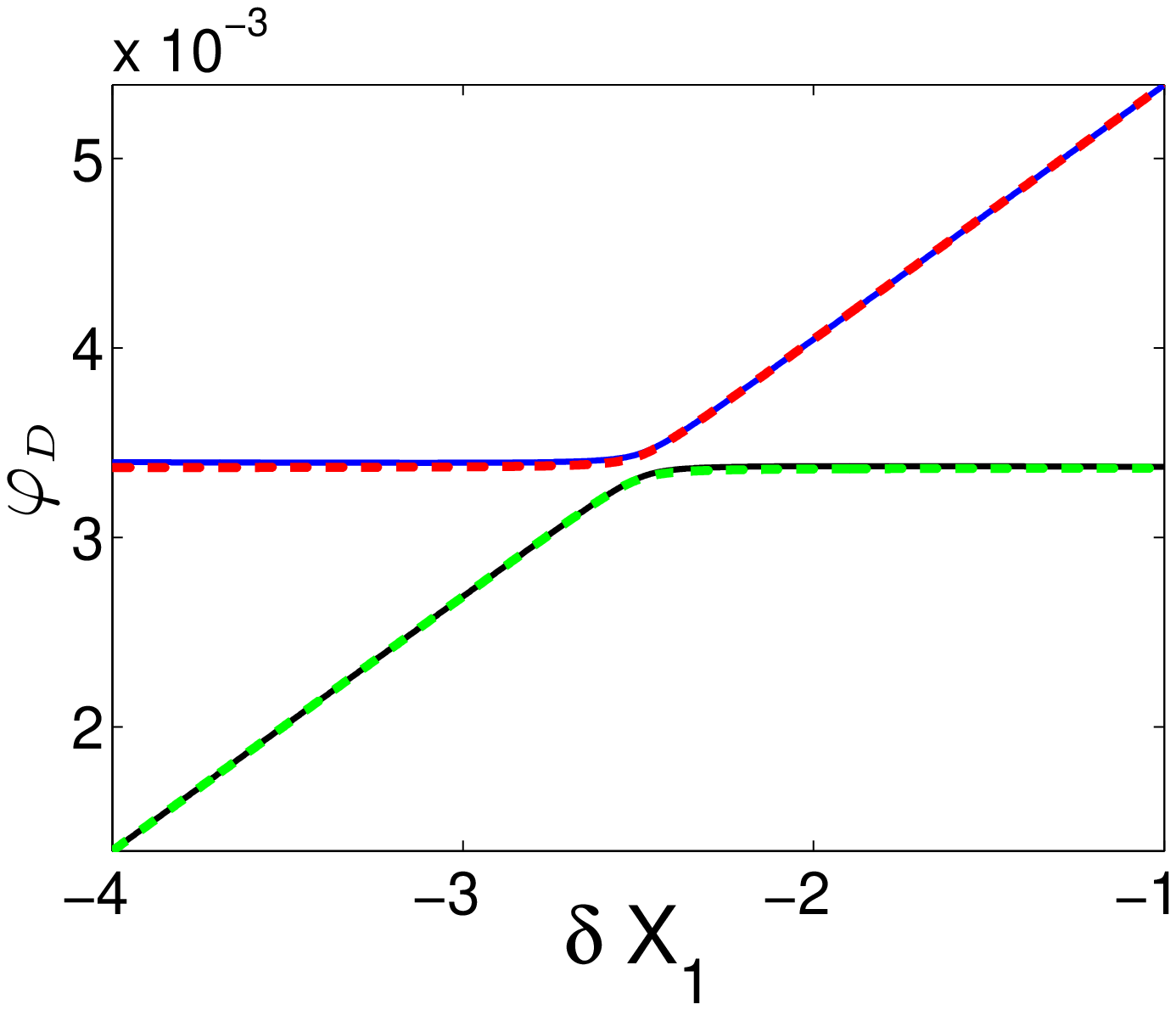}

{\footnotesize {\bf FIG. 7.} 
Tests of the perturbation theory based 
approximations (dashed lines) against 
the numerics (solid lines).
The model parameters are ${L_{\tbox{D}}=\mass=e=1}$   
and ${L_{\tbox{W}} = 3000.43}$, and we focus 
on the degeneracy point $n_r = 2993$.
For these parameters ${X^{(r)} \approx 465}$ 
and ${g^{(r)} \approx 4.5 \times 10^{-5}}$.
All the plots refer to the path $(X_2 - X_1) = 5$.
In the left panel the dashed lines are derived 
from Eq.(\ref{e45}). In the middle panel the dashed 
lines are based on Eqs.(\ref{e48}-\ref{e49}) 
with $\theta$ from Eq.(\ref{e44}).
In the right panel the dashed lines are deduced 
from Eq.(\ref{e38})}

}

\ \\ \ \\

\mpg{


\putgraph[width=0.49\hsize]{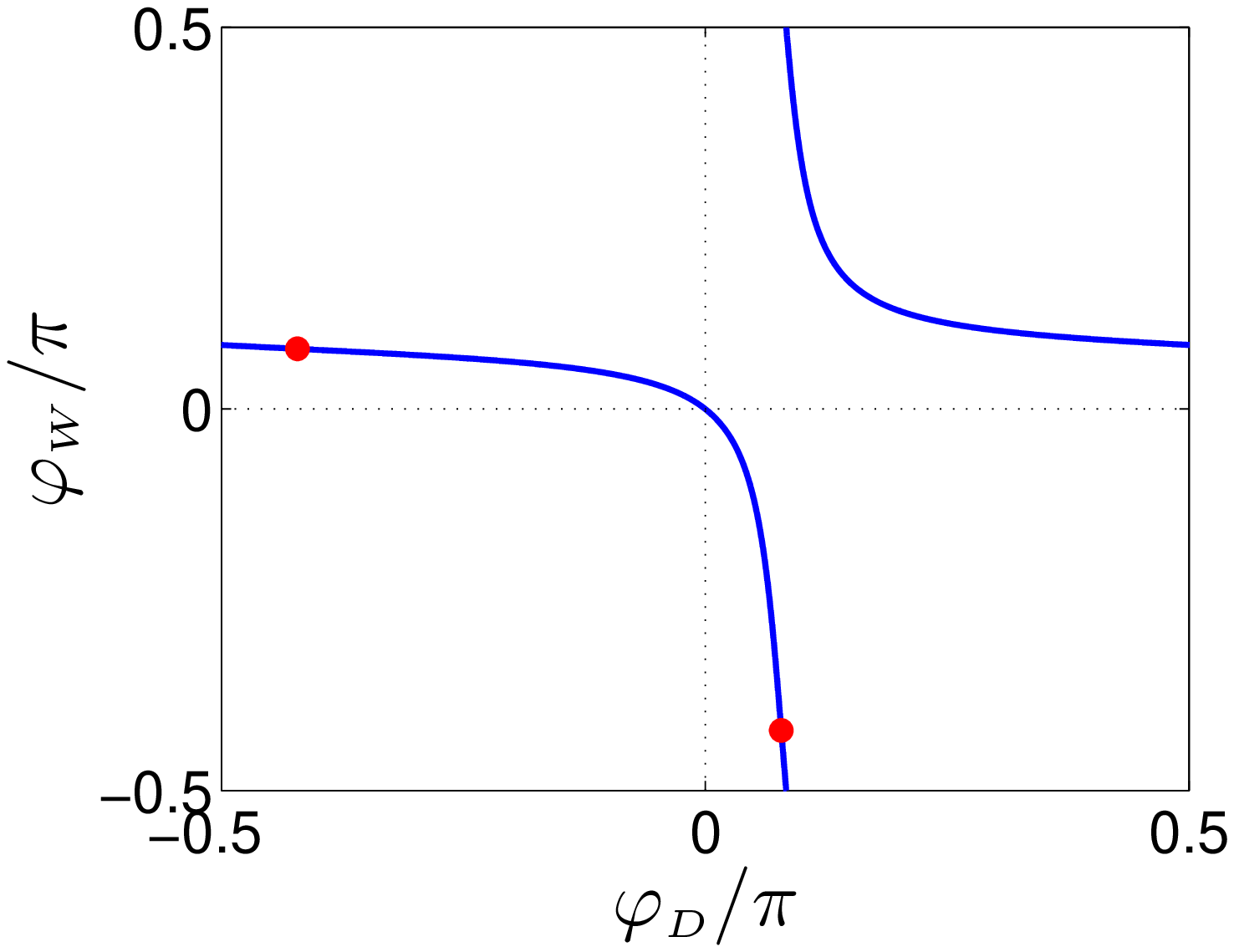}

{\footnotesize {\bf FIG. 8.} }
The wire phase $\varphi_{\tbox{W}}/\pi$ versus 
the dot phase $\varphi_{\tbox{D}}/\pi$ 
at a node with delta barrier $g(X) = 0.225$.  
The two branches are implied by the 
matching condition Eq.(\ref{e27}). The ratio 
$|{\sin(\varphi_{\tbox{D}})}/{\sin(\varphi_{\tbox{D}})}|$ 
attains its extremal values (Eq.(\ref{e35})) 
at the points which are indicated by circles.}


\Bibliography{99}

\bibitem{rev}
L.P. Kouwenhoven, C.M. Marcus, P.L. Mceuen,
S. Tarucha, R. M. Westervelt and N.S. Wingreen,
Proc. of Advanced Study Inst. on Mesoscopic
Electron Transport, edited by L.L. Sohn,
L.P. Kouwenhoven and G. Schon (Kluwer 1997).



\bibitem{pmp}
D. Cohen, cond-mat/0208233 (2002); 
Solid State Communications {\bf 133}, 583 (2005).

\bibitem{pmc}
D. Cohen, Phys. Rev. B {\bf 68}, 155303 (2003).

\bibitem{pme}
For a mini-review see: D. Cohen, Physica E 28, 308 (2005).

\bibitem{pml} 
G.~Rosenberg and D.~Cohen, 
cond-mat/0510289, J. Phys. A (2006, in press).




\bibitem{tbm}
Y. Levinson, O. Entin-Wohlman and P. Wolfle, cond-mat/0010494.

\bibitem{BPT1}
M. Buttiker, H. Thomas and A Pretre, 
Z.~Phys.~B-Condens.~Mat., {\bf 94}, 133 (1994).  

\bibitem{BPT2}
P. W. Brouwer, Phys. Rev. B {\bf 58}, R10135 (1998).

\bibitem{SAA}
T. A. Shutenko, I. L. Aleiner and B. L. Altshuler, 
Phys. Rev. {\bf B61}, 10366 (2000).

\bibitem{pMB}
M.~Moskalets and M.~B{\"u}ttiker, 
Phys. Rev. B {\bf 68}, 161311(R) (2003).




\bibitem{berry1}
M.V. Berry, Proc. R. Soc. Lond. A {\bf 392}, 45 (1984).

\bibitem{thouless}
D. J. Thouless, 
Phys. Rev. B {\bf 27}, 6083 (1983).

\bibitem{AvronNet}
J.E. Avron, A. Raveh and B. Zur,  
Rev. Mod. Phys. {\bf 60}, 873 (1988).

\bibitem{berry2}
M.V. Berry and J.M. Robbins, Proc. R. Soc. Lond. A {\bf 442}, 659 (1993).

\bibitem{pmm} 
I.~Sela and D.~Cohen, in preparation. 

\bibitem{pmo}
D. Cohen, Phys. Rev. B {\bf 68}, 201303(R) (2003).


 
\end{thebibliography}
\end{document}